\documentclass[10pt,twocolumn]{IEEEtran}
\usepackage{amsmath}
\usepackage{amsthm,amssymb,amsmath,bm}
\usepackage{subfigure}
\usepackage{amsfonts}
\usepackage{epsfig}
\usepackage{algcompatible}
\usepackage{amssymb}
\usepackage{color}
\usepackage{transparent}
\usepackage{graphicx}
\usepackage{amsmath}
\graphicspath{{image/}}
\usepackage{epstopdf}
\usepackage[table,xcdraw]{xcolor}
\usepackage{cite}
\usepackage[utf8x]{inputenc}
\usepackage{color,soul}
\usepackage{graphicx}
\usepackage{epstopdf}
\usepackage{graphics}
\usepackage{subfigure}
\usepackage{multirow}
\usepackage{rotating}
\usepackage{graphicx}
\usepackage{tabularx}
\usepackage{array}
\usepackage{color,soul}
\usepackage{bm}
\usepackage{graphicx,dblfloatfix}
\usepackage{blindtext}

\usepackage{subfigure}
\usepackage{moreverb}
\usepackage{epsfig}
\usepackage{amsmath,amssymb,amsthm,mathrsfs,amsfonts,dsfont}
\usepackage{epstopdf}
\usepackage{adjustbox,lipsum}
\usepackage{amsfonts}
\usepackage{epsfig}
\usepackage{amssymb}
\usepackage{amsmath}
\usepackage{amsthm}
\usepackage{subfigure}
\usepackage{multirow}
\usepackage{rotating}
\usepackage{graphicx}
\usepackage{tabularx}
\usepackage{array}
\usepackage{anyfontsize}
\usepackage{color,soul}
\usepackage{graphicx,dblfloatfix}
\usepackage{epstopdf}
\usepackage{blindtext}
\usepackage{amsmath}
\usepackage{amsthm,amssymb,amsmath,bm}
\usepackage{subfigure}
\usepackage{amsfonts}
\usepackage{epsfig}
\usepackage{amssymb}
\usepackage{amsfonts}
\usepackage{amsmath}
\usepackage{cite}
\usepackage{hyperref}
\usepackage{graphicx}
\usepackage{fancyhdr}
\usepackage{subfigure}
\usepackage[subfigure]{tocloft}
\usepackage[font={small}]{caption}
\usepackage{subfigure}
\usepackage{tabularx}
\usepackage{tcolorbox}
\usepackage{cite}
\usepackage[linesnumbered,ruled,vlined]{algorithm2e}
\SetKwInput{KwInput}{Input}
\SetKwInput{KwOutput}{Output}
\usepackage{amsthm,amssymb,amsmath,bm}
\usepackage{graphicx}
\usepackage{fancyhdr}
\usepackage{subfigure}
\usepackage[subfigure]{tocloft}
\usepackage[font={small}]{caption}
\usepackage{subfigure}
\usepackage{tabularx}
\usepackage{cite}
\allowdisplaybreaks
\usepackage{lettrine}
\usepackage{lipsum,parskip,libertine}
\usepackage{blindtext}
\usepackage{slashbox}
\usepackage{diagbox}

\IEEEdisplaynontitleabstractindextext
\IEEEpeerreviewmaketitle
\begin{document}
	\title{AI-based Robust Resource Allocation in End-to-End Network Slicing under Demand and CSI Uncertainties}
	\author{\IEEEauthorblockN{Amir Gharehgoli, Ali Nouruzi, \IEEEmembership{Student Member, IEEE}, Nader Mokari, Paeiz Azmi, Mohamad Reza Javan, \IEEEmembership{Senior Member, IEEE}, Eduard A. Jorswieck, \IEEEmembership{Fellow Member, IEEE}
			\thanks{
				A. Gharehgoli, A. Nouruzi, N. Mokari, and P. Azmi are with the Department of Electrical and Computer Engineering,~Tarbiat Modares University,~Tehran,~Iran, (e-mail:\{a.gharahgoli, ali\_nouruzi, nader.mokari, pazmi\}@modares.ac.ir). M.~R.~Javan is with the Department of Electrical Engineering, Shahrood University of Technology, Iran, (e-mail: javan@shahroodut.ac.ir). Eduard A. Jorswieck is with TU Braunschweig, Department of Information Theory and Communication Systems, Braunschweig,
				Germany (jorswieck@ifn.ing.tu-bs.de).
			}}}
		\maketitle
\begin{abstract}
Network slicing (NwS) is one of the main technologies in the fifth-generation of mobile communication and beyond (5G+). One of the important challenges in the NwS is information uncertainty which mainly involves demand and channel state information (CSI). Demand uncertainty is divided into three types: number of users requests, amount of bandwidth, and requested virtual network functions workloads. Moreover, the CSI uncertainty is modeled by three methods: worst-case, probabilistic, and hybrid. In this paper, our goal is to maximize the utility of the infrastructure provider by exploiting deep reinforcement learning algorithms in end-to-end NwS resource allocation under demand and CSI uncertainties. The proposed formulation is a non-convex mixed-integer non-linear programming problem. To perform robust resource allocation in problems that involve uncertainty, we need a history of previous information. To this end, we use a recurrent deterministic policy gradient (RDPG) algorithm, a recurrent and memory-based approach in deep reinforcement learning. Then, we compare the RDPG method in different scenarios with soft actor-critic (SAC), deep deterministic policy gradient (DDPG), distributed, and greedy algorithms. The simulation results show that the SAC method is better than the DDPG, distributed, and greedy methods, respectively. Moreover, the RDPG method out performs the SAC approach on average by 70$\%$.
	\end{abstract}
\section{Introduction}
\lettrine{M}{obile} devices are becoming a necessary part of our everyday life, and with the increase of wireless devices, mobile data traffic is growing exponentially \cite{union2015imt,foukas2017network,habiba2018auction,debbabi2020algorithmics}. To this end, with the advent of the fifth-generation of mobile communication (5G) in recent years, a wide variety of services has emerged. In other words, 5G provides heterogeneous services customized for users based on their specific needs. A physical network is split into several dedicated logical networks by network slicing (NwS) to satisfy the requirements of various use cases \cite{zhou2020dynamic}. By deploying NwS, mobile network operators are able to split the physical infrastructure into isolated virtual networks (slices), which are managed by service providers to provide customized services. The concept of NwS allows infrastructure providers (InPs) to provide heterogeneous 5G services over a common platform. Slice requests are accepted by the InP to generate revenue. The 3rd generation partnership project (3GPP) \cite{dogra2020survey} and international telecommunication union (ITU) \cite{habibi2019comprehensive} have divided all 5G slices into three categories: enhanced mobile broadband (eMBB), ultra-reliable low latency communications (uRLLC), and massive machine-type communication (mMTC). Each type of slice has specific requirements. The eMBB slice needs a high data rate and includes services such as web browsing, video streaming, virtual reality, etc. The uRLLC slice requires high reliability and low latency that consist of services like cloud gaming, remote surgery, autonomous driving, etc. The mMTC slice needs efficient connectivity for a massive number of devices, e.g., sensor networks. The concept of end-to-end (E2E) means that the resources of the radio access network (RAN) and the core network are considered together simultaneously. To maintain the quality of service (QoS) and ensure the E2E delay of these slices, we need to apply new technologies and strategies. In 5G, software-defined networking (SDN) and network function virtualization (NFV) fill the void of programmable control and management of network resources. Network management is easier with SDN since it decouples the control plane from the data plane while centralizing the network's intelligence. NFV enables the implementation of originally hardware-based proprietary network functions on virtual environments. The virtual network functions (VNFs) are run on virtual machines (VMs) or containers to provide network or value-added services. Note that they are chained together in a co-located or distributed cloud environment \cite{afolabi2018network}. SDN and NFV cause a significant increase in the cellular network's programmability, agility, scalability, flexibility, and development that can reduce the capital expense (CAPEX) and operation expense (OPEX) \cite{nguyen2017sdn,ordonez2017network}. Every service consists of a predefined sequence of functions, called service function chains (SFCs). 
\\In recent years, deploying machine learning (ML) algorithms to address the challenges of the cellular network, such as information uncertainty (for example, channel state information (CSI) and demand), has significantly increased. Therefore, in this work to robust resource allocation and address uncertainty issues, we exploit this algorithms to solve our system model.
\subsection{Paper Organization}
The rest of this paper is organized as follows. In Section \ref{II}, related works are summarized. Section \ref{III} explains the system model. Section \ref{IV} introduces our problem formulation. The several deep reinforcement learning (DRL) algorithms to solve the proposed problem are developed in Section \ref{V}. We provide Section \ref{VI} to understand the computational complexity and convergence analysis of the solutions better. The simulation results under various conditions are presented in Section \ref{VII}. Finally, the conclusions of this paper are discussed in Section \ref{VIII}.\\
\textbf{Symbol Notations:}
To denote the vector A and the element $i$-th of this vector, we use the bold upper symbol as $\bold{A}$ and $a_i$, respectively. Likewise, $a_{i,j}$ define  the of  matrix $\bold{A}$. In addition, $\mathcal{B}$ and $b_j$ denote the set B and the $j$-th element of it, respectively. We use $|c|$ to define absolute value of $c$. To define the expectation of $d$, we use $\mathbb{E}[d]$. Moreover, we use $\text{Var}[e]$ to determine the variance $e$.
\section{Related Works}\label{II}
The purpose of this section is to review the related works and categorize them based on our contribution. To this end, we divide the related works into five categories: i) Resource allocation problems in RAN slicing \cite{zhou2020radio,tang2019service,d2018low,korrai2020ran,sun2020efficient,wang2019reinforcement,lee2018dynamic}, ii) Resource allocation problems in core slicing \cite{ebrahimi2020joint,chen2020network,promwongsa2020ensuring}, iii) Resource allocation problems in E2E slicing \cite{li2020end,tong2020communication,lin2018end,harutyunyan2019orchestrating,liu2020provisioning,mei20205g}, iv) NwS problems under demand uncertainty \cite{esmat2020deep,reddy2016robust,baumgartner2017network,kasgari2018stochastic,wen2018robustness,van2019optimal,khamse2021agile}, v) NwS problems under CSI uncertainty \cite{jumba2015energy,khan2020deep,korrai2020joint,moltafet2019robust}. Each of the categories is described below. Moreover, we summarize the related works and compare them with our work in Table. \ref{Summary of related works}.
\subsection{Resource Allocation Problems in RAN Slicing}
A radio resource allocation method to maximize the SLA contract rate and maintain the isolation between the slices is introduced in \cite{zhou2020radio}. The problem is solved using the Lagrange dual algorithm. The authors in \cite{tang2019service} introduce the cloud RAN (C-RAN) operator’s revenue maximization by correctly accepting the slice requests. Two types of long-term and short-term revenues are considered. The long-term revenue is calculated by the values in the network slice request, and the short-term revenue is achieved by saving system power consumption in each frame. The optimization problem is formulated as a mixed-integer nonlinear programming (MINP), then to solve the problem, the authors employ a successive convex approximation (SCA) algorithm. \cite {d2018low} proposes the near-optimal low-complexity distributed RAN slicing problem as a congestion game. The problem aims to minimize the cost. A heuristic approach to maximize sum-rate in a single-cell cellular network scenario is presented in \cite{korrai2020ran}. The authors formulate a resource allocation problem subject to service isolation, latency, and minimum rate constraints. To maintain the reliability constraint, they use adaptive modulation and coding. \cite{sun2020efficient} addresses the smart handover mechanism by employing a multi-agent Q-learning to minimize handover cost while guaranteeing a various QoS requirements. To compute the cost, the authors define four handover cost types: i) the cost of switching service types when user equipment (UE) remains in the coverage range of the same base station (BS). ii) handover cost when a UE leaves the coverage of a BS with the same service type. iii) the cost associated with user movement and changing service types. iv) the cost of implementing a new network slice to maintain the QoS of the user handover. \cite{wang2019reinforcement} introduce the NwS resource allocation problem in 5G C-RAN to maximize operators' utility. The framework of the problem includes an upper layer that manages the mapping of virtual protocol stack functions; and a lower layer, which controls radio remote unit association, power, and subchannel allocation. To reduce the complexity of the Q-value table, the authors used the multi-agent Q-learning approach. In \cite{lee2018dynamic}, the authors investigate a dynamic NwS framework for downlink multi-tenant heterogeneous cloud RAN (H-CRAN) by considering both small-cell and macro-cell tiers. The proposed architecture includes two-level, an upper level for managing admission control, baseband resource allocation, and user association, and a lower level for handling radio resource allocation between users. The objective of the problem is to maximize the throughput of the tenants by considering the constraints of QoS, fronthaul and backhaul capacities, tenants' priorities, baseband resources, and interference.
\subsection{Resource Allocation Problems in Core Slicing}
In \cite{ebrahimi2020joint}, the authors introduced a novel joint admission and resource management approach in multiple tenants scenario to minimize the cost of bandwidth consumption and power consumption cost of all turned-on cloud nodes. The main objective of \cite{chen2020network} is to minimize the total power consumption of a cloud node, which consists of the static power consumption and the dynamic load-dependent power consumption. The authors consider the resource budget, functional instantiation, flow routing, and guarantee the E2E latency of all services, where E2E delay consists of total NFV delay on the cloud nodes and total communication delay on the links. The problem is formulated as a mixed binary linear program, then it is solved by a heuristic approach. \cite{promwongsa2020ensuring} introduces the VNF placement problem in NwS for serving tactile applications and routing tactile traffic.  The main goal of the optimization problem is to minimize reliability degradation in addition to maintaining the strict delay constraint. To find sub-optimal solutions, a Tabu search-based algorithm is used due to the complexity of the formulated problem. 
\\To the best of our knowledge, the objective function of most of the papers in core slicing related to energy consumption and are solved mainly using heuristic methods.
\begin{table*}[h!]
	\centering
	\caption{Summary of the related works}
	\label{Summary of related works}
	\scalebox{0.88}{
		\begin{tabular}{|c|c|c|c|c|c|}
			\hline
					\rowcolor[HTML]{38FFF8} 
			\textbf{Ref.}                                        & \textbf{Objective Function}                            & \textbf{\begin{tabular}[c]{@{}c@{}}Slicing \\ Domain\end{tabular}} & \textbf{Optimization Algorithm}                 & \textbf{\begin{tabular}[c]{@{}c@{}}Demand\\ Uncertainty\end{tabular}} & \textbf{\begin{tabular}[c]{@{}c@{}}CSI\\ Uncertainty\end{tabular}} \\ \hline
			\cite{zhou2020radio}                & SLA contract rate maximization                         & RAN                                                                & Lagrangian dual                                 & $\color{red}{\pmb{\mathsf{x}}}$                                       & $\color{red}{\pmb{\mathsf{x}}}$                                    \\ \hline
			\cite{tang2019service}              & Revenue maximization                                   & C-RAN                                                              & SCA                 & $\color{red}{\pmb{\mathsf{x}}}$                                       & $\color{red}{\pmb{\mathsf{x}}}$                                    \\ \hline
			\cite{d2018low}                     & Cost minimization                                      & RAN                                                                & Game theory                                     & $\color{red}{\pmb{\mathsf{x}}}$                                       & $\color{red}{\pmb{\mathsf{x}}}$                                    \\ \hline
			\cite{korrai2020ran}                & Sum-rate maximization                                  & RAN                                                                & Heuristic                                       & $\color{red}{\pmb{\mathsf{x}}}$                                       & $\color{red}{\pmb{\mathsf{x}}}$                                    \\ \hline
			\cite{sun2020efficient}             & Handover cost minimization                             & RAN                                                                & Multi-agent Q-learning                          & $\color{red}{\pmb{\mathsf{x}}}$                                       & $\color{red}{\pmb{\mathsf{x}}}$                                    \\ \hline
			\cite{wang2019reinforcement}        & Utility maximization                                   & C-RAN                                                              & Multi-agent Q-learning                          & $\color{red}{\pmb{\mathsf{x}}}$                                       & $\color{red}{\pmb{\mathsf{x}}}$                                    \\ \hline
			\cite{lee2018dynamic}               & Throughput maximization                                & H-CRAN                                                             & Greedy \& Lagrangian dual                       & $\color{red}{\pmb{\mathsf{x}}}$                                       & $\color{red}{\pmb{\mathsf{x}}}$                                    \\ \hline
			\cite{ebrahimi2020joint}            & Cost minimization                                      & Core                                                               & Heuristic                                       & $\color{red}{\pmb{\mathsf{x}}}$                                       & $\color{red}{\pmb{\mathsf{x}}}$                                    \\ \hline
			\cite{chen2020network}              & Energy consumption minimization                        & Core                                                               & Heuristic                                       & $\color{red}{\pmb{\mathsf{x}}}$                                       & $\color{red}{\pmb{\mathsf{x}}}$                                    \\ \hline
			\cite{promwongsa2020ensuring}       & Cost minimization                                      & Core                                                               & Tabu search                                     & $\color{red}{\pmb{\mathsf{x}}}$                                       & $\color{red}{\pmb{\mathsf{x}}}$                                    \\ \hline
			\cite{li2020end}                    & E2E access rate maximization                           & E2E                                                                & Deep Q-Learning                                 & $\color{red}{\pmb{\mathsf{x}}}$                                       & $\color{red}{\pmb{\mathsf{x}}}$                                    \\ \hline
			\cite{tong2020communication}        & Delay minimization                                     & E2E                                                                & VF2                                             & $\color{red}{\pmb{\mathsf{x}}}$                                       & $\color{red}{\pmb{\mathsf{x}}}$                                    \\ \hline
			\cite{lin2018end}                   & Utility maximization                                   & E2E                                                                & Primal-dual Newton                              & $\color{red}{\pmb{\mathsf{x}}}$                                       & $\color{red}{\pmb{\mathsf{x}}}$                                    \\ \hline
			\cite{harutyunyan2019orchestrating} & Number of VNF migrations minimization                  & E2E                                                                & Heuristic                                       & $\color{red}{\pmb{\mathsf{x}}}$                                       & $\color{red}{\pmb{\mathsf{x}}}$                                    \\ \hline
			\cite{ liu2020provisioning}         & QoS maximization \& resource  consumption minimization & E2E                                                                & Heuristic                                       & $\color{red}{\pmb{\mathsf{x}}}$                                       & $\color{red}{\pmb{\mathsf{x}}}$                                    \\ \hline
			\cite{mei20205g}                    & Resource consumption minimization                      & E2E                                                                & Heuristic                                       & $\color{red}{\pmb{\mathsf{x}}}$                                       & $\color{red}{\pmb{\mathsf{x}}}$                                    \\ \hline
			\cite{esmat2020deep}                & Revenue maximization                                   & E2E                                                                & Q-learning                                      & $\color{green}{\checkmark}$                                           & $\color{red}{\pmb{\mathsf{x}}}$                                    \\ \hline
			\cite{reddy2016robust}              & Cost minimization                                      & Core                                                               & Heuristic-based on $\Gamma$-robustness                 & $\color{green}{\checkmark}$                                           & $\color{red}{\pmb{\mathsf{x}}}$                                    \\ \hline
			\cite{baumgartner2017network}       & Cost minimization                                      & Core                                                               & Light robustness-based on $\Gamma$-robustness          & $\color{green}{\checkmark}$                                           & $\color{red}{\pmb{\mathsf{x}}}$                                    \\ \hline
			\cite{kasgari2018stochastic}        & Power minimization                                     & RAN                                                                & Heuristic-based on Lyapunov drift plus-penalty  & $\color{green}{\checkmark}$                                           & $\color{red}{\pmb{\mathsf{x}}}$                                    \\ \hline
			\cite{wen2018robustness}            & Bandwidth consumption minimization                     & E2E                                                                & Heuristic-based on variable neighborhood search & $\color{green}{\checkmark}$                                           & $\color{red}{\pmb{\mathsf{x}}}$                                    \\ \hline
			\cite{van2019optimal}               & Long-term return maximization                          & E2E                                                                &Double deep  Q-learning                          & $\color{green}{\checkmark}$                                           & $\color{red}{\pmb{\mathsf{x}}}$                                    \\ \hline
			\cite{khamse2021agile}              & Utility maximization                                   & E2E                                                                & Iterative auction game                          & $\color{green}{\checkmark}$                                           & $\color{red}{\pmb{\mathsf{x}}}$                                    \\ \hline
			\cite{jumba2015energy}             & Energy efficiency maximization                         & RAN                                                                & Iterative-based on Lagrangian dual              & $\color{red}{\pmb{\mathsf{x}}}$                                       & $\color{green}{\checkmark}$                                        \\ \hline
			\cite{khan2020deep}                 & Threshold rate violation probability minimization      & RAN                                                                & Deep neural network                             & $\color{red}{\pmb{\mathsf{x}}}$                                       & $\color{green}{\checkmark}$                                        \\ \hline
			\cite{korrai2020joint}              & Power minimization                                     & RAN                                                                & SCA                 & $\color{red}{\pmb{\mathsf{x}}}$                                       & $\color{green}{\checkmark}$                                        \\ \hline
			\cite{moltafet2019robust}           & Sum-rate maximization                                  & C-RAN                                                              & SCA                 & $\color{red}{\pmb{\mathsf{x}}}$                                       & $\color{green}{\checkmark}$                                        \\ \hline
			Our work                                             & Utility maximization                                   & E2E                                                                & Recurrent deterministic policy gradient                                            & $\color{green}{\checkmark}$                                           & $\color{green}{\checkmark}$                                        \\ \hline
		\end{tabular}
	}
\end{table*}
\subsection{Resource Allocation Problems in E2E Slicing}
Based on the deep Q-networks (DQN) algorithm, \cite{li2020end} investigates the dynamic resource allocation problem to maximize the E2E access rate for multi-slice and multi-service scenarios. The purpose of \cite{tong2020communication} is to minimize the E2E delay to guarantee the reliability requirements of NwS in the uRLLC application. Based on subgraph isomorphism, the authors propose a communication and computation resource allocation algorithm. For multi-service converged infrastructure, an iterative primal-dual fast network resource slicing algorithm is formulated in \cite{lin2018end}. The main aim is to maximize the total utility and satisfy system constraints while jointly optimizing flow routing, power slicing, and congestion control. To solve the proposed optimization problem, the authors employ the primal-dual Newton’s approach. A mixed-integer linear programming (MILP) slice placement problem is studied in \cite{harutyunyan2019orchestrating} to compare and examine different E2E slice placement approaches. The SFC is used to model E2E slice requests in which each RAN and core network component are represented as a VNF. The problem is about optimizing network utilization and ensuring that the QoS requirements of the considered slice requests are met by the objective of minimizing the number of VNF migrations in the network and their impacts on the slices. In \cite{liu2020provisioning} a hierarchical information-centric networking system without requiring prior knowledge on the VN’s topology and resource provisioning information is studied, then a heuristic algorithm is investigated to solve the integer linear program (ILP) problem. The proposed resource allocation problem aims to find the best trade-off between QoS and resource consumption, using VNFs and link bandwidth as resources. The E2E shareable-VNFs-based multiple couple virtual network embedding (VNE) problem is addressed in \cite{mei20205g}. To minimize the physical resources consumption, the sharing property of VNFs is considered. The problem is modeled as ILP, and is solved by a heuristic algorithm. The authors categorize VNF types into shareable and non-shareable ones, then compute the different resource requirements of them. Moreover, they show that the slice acceptance ratio on the same physical network using VNF-sharing can be improved.
\subsection{NwS Problems under Demand Uncertainty}
A new dynamic edge/fog NwS (EFNwS) scheme to find an optimal slice request admission policy to maximize the InP's long-term revenue is proposed in \cite{esmat2020deep}. Tenants can temporarily lease back to the InP the unused resources to serve demands exceeding its current resources in stock. A semi-Markov decision process (SMDP) is used to model the arrival of slice requests. A Q-learning algorithm is applied to find the optimal policy under uncertain resource demands. Moreover, to reduce the computational complexity of Q-learning and improve the convergence time, a DRL algorithm and an enhancement based on a deep dueling (Dueling DQ-EFNwS) algorithm are applied. In \cite{reddy2016robust}, a novel optimization model based on the concept of $\Gamma$-robustness to deal with uncertainty in the traffic demand is proposed. The $\Gamma$-robust optimization is formulated as a MILP.  A modified MIP-based variable neighborhood search (VNS) heuristic is provided to enhance the model's scalability. The authors in \cite{baumgartner2017network} introduce a novel model applying the concept of light robustness to address scalability issues of traffic uncertainty in NwS and to get a deeper insight into the trade-off between the price of robustness and the realized robustness. NwS in the wireless system with a time-varying number of users that require two types of slices: reliable low latency (RLL) and self-managed (capacity limited) slices are studied in \cite{kasgari2018stochastic}. A novel control framework for stochastic optimization based on the Lyapunov drift-plus-penalty method is proposed. This framework enables the system to minimize power, maintain slice isolation, and provide reliable and low latency E2E communication for RLL slices. Robust NwS by addressing the slice recovery and reconfiguration with stochastic traffic demands in each slice is introduced in \cite{wen2018robustness}. For solving the optimization problem, a heuristic algorithm based on VNS is developed. \cite{van2019optimal} provides an optimal and fast resource slicing solution under the uncertainty of resource demand from tenants that maximizes the long-term return of the network provider. An SMDP is employed to allocate resources to users under the dynamic demands of users. A novel approach using an advanced deep Q-learning technique called the deep dueling algorithm is adopted to obtain the optimal resource allocation policy for the network provider. The distributed online approach for inter-domain resource allocation to network slices in a heterogeneous multi-resource multi-domain mobile network environment is investigated in \cite{khamse2021agile}. The main goal of the work is to maximize the utility of network slice instances while minimizing the OPEX for infrastructure service providers; at the same time, then; the iterative auction game among network slice tenants is employed to solve the proposed problem.
\subsection{NwS Problems under CSI Uncertainty}
In \cite{jumba2015energy}, the authors propose a robust resource allocation in virtualized wireless networks (VWNs) to address the uncertainty in CSI at the BS. Based on a newly defined slice utility function, a robust resource allocation problem is formulated, aiming to maximize the overall energy efficiency of VWN under the worst-case CSI uncertainty. Uncertain CSI is defined as the sum of its true estimated value, and an error is considered to be limited in a specific uncertainty rang. A dynamic resource allocation problem for vehicular UE requesting eMBB and uRLLC slices is proposed in \cite{khan2020deep}. The main objective of the problem is the minimum threshold rate violation probability for the eMBB slice while it guarantees a probabilistic threshold rate for the uRLLC slice. A deep neural network (DNN) is used to estimate CSI based on the propagation environment such as scatterers and reflectors. In \cite{korrai2020joint}, the authors investigate the joint power and resource blocks allocation to the heterogeneous users by considering mixed numerology-based frame structures. In addition to considering each service's heterogeneous QoS requirements, the authors consider each user's queue condition when scheduling resource blocks. The objective function of the proposed problem is to minimize the overall power consumption of the BS for each sub-frame. Outage probabilistic constraints are included to deal with imperfect CSI. \cite{moltafet2019robust} presents a novel robust radio resource allocation under worst-case CSI uncertainty in downlink channel of a sparse code multiple access (SCMA) based C-RAN by considering multiple-input and single-output transmission technology. The main objective of the suggested problem is to maximize the sum-rate under several conditions such as user association, the minimum rate required of each slice, maximum available power at radio remote head (RRH), maximum fronthaul capacity of each RRH, and SCMA constraints. A two-step iterative algorithm based on SCA is applied to solve the formulated optimization problem.
\\According to the references presented in Section \ref{II}, most studies in NwS mainly considered resources of RAN or core \cite{zhou2020radio,tang2019service,d2018low,korrai2020ran,sun2020efficient,wang2019reinforcement,lee2018dynamic,ebrahimi2020joint,chen2020network,promwongsa2020ensuring}, so there are not many E2E NwS papers and usually they do not include demand or CSI uncertainties \cite{li2020end,tong2020communication,lin2018end,harutyunyan2019orchestrating,liu2020provisioning,mei20205g}. In addition, studies that include uncertainty address only one type of uncertainty \cite{esmat2020deep,reddy2016robust,baumgartner2017network,kasgari2018stochastic,wen2018robustness,van2019optimal,khamse2021agile,jumba2015energy,khan2020deep,korrai2020joint,moltafet2019robust}. To the best of our knowledge, none of the previous studies meet all the conditions set out in our work.
\subsection{Contributions}
The main contributions of this paper are summarized as follows:
\\$\bullet$ The resource allocation and management in NwS must be E2E. The data rate requested by users in each slice from InP may be stochastic. Moreover, the CSI in the BS due to different reasons such as mobility of users is imperfect. To this end, in this paper,  to bring the proposed problem closer to the real scenario, we formulate a resource allocation problem in E2E NwS under both demand and CSI uncertainties. 
\\$\bullet$ The robust mathematical optimization methods and ML algorithms are two methods that are commonly used to solve problems that include uncertainty. We need to consider the history of previous information. Therefore we employ the recurrent deterministic policy gradient (RDPG) algorithm as the main approach. Then, we compare it with soft actor-critic (SAC), deep deterministic policy gradient (DDPG), distributed, and greedy algorithms from various aspects.
\\$\bullet$ Simulation results show that the SAC strategy outperform the DDPG, distributed, and greedy algorithms, respectively. Additionally, the RDPG approach also outperforms the SAC method by approximately 70$\%$. As a result, we consider the RDPG approach the most suitable and robust method for our proposed problem.
\section{System Model}\label{III}
\begin{figure}[h!]
	\centering
	\includegraphics[width=0.94\linewidth]{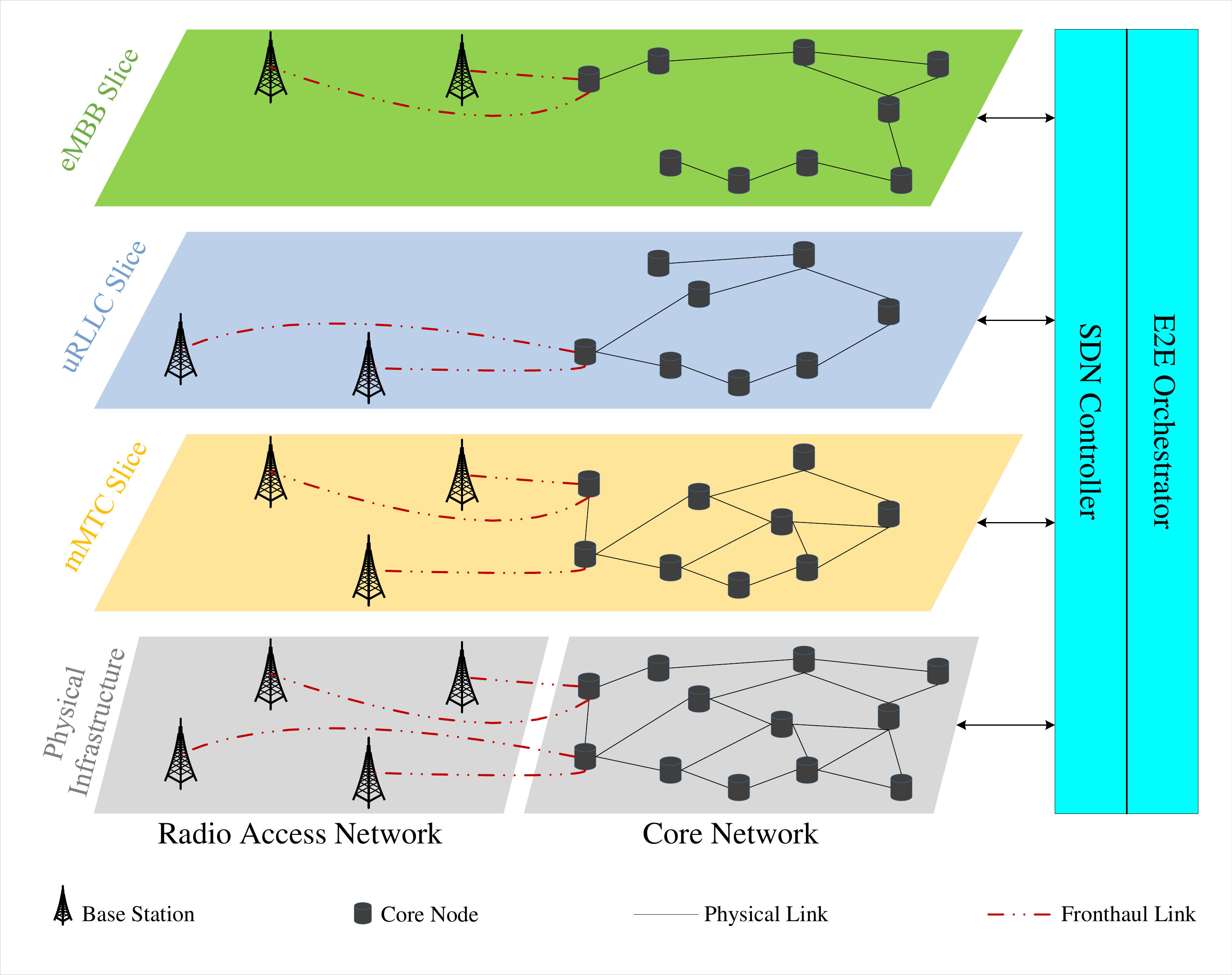}
	\caption{Illustration of the E2E NwS structure.}
	\label{Fig E2E}
\end{figure}

This section describes our NwS architecture and system model. In this paper, we assume the E2E NwS with a single InP and multiple users. The users request slices from the InP. We consider the SDN controller to manage the network and improve performance. The system model consists of two parts; the RAN domain and the core domain. By backhaul links, these two parts are connected through the transport domain. The proposed E2E NwS is depicted in Fig. \ref{Fig E2E}. In Subsections, \ref{RAN} and \ref{Core} the details of the RAN and core network of our system model are described, respectively. Moreover, to increase the readability of this paper, the main parameters and variables are listed in Table \ref{Main notations}.

\subsection{Radio Access Network Explanation}\label{RAN}
\begin{figure}[h!]
	\centering
	\includegraphics[width=0.94\linewidth]{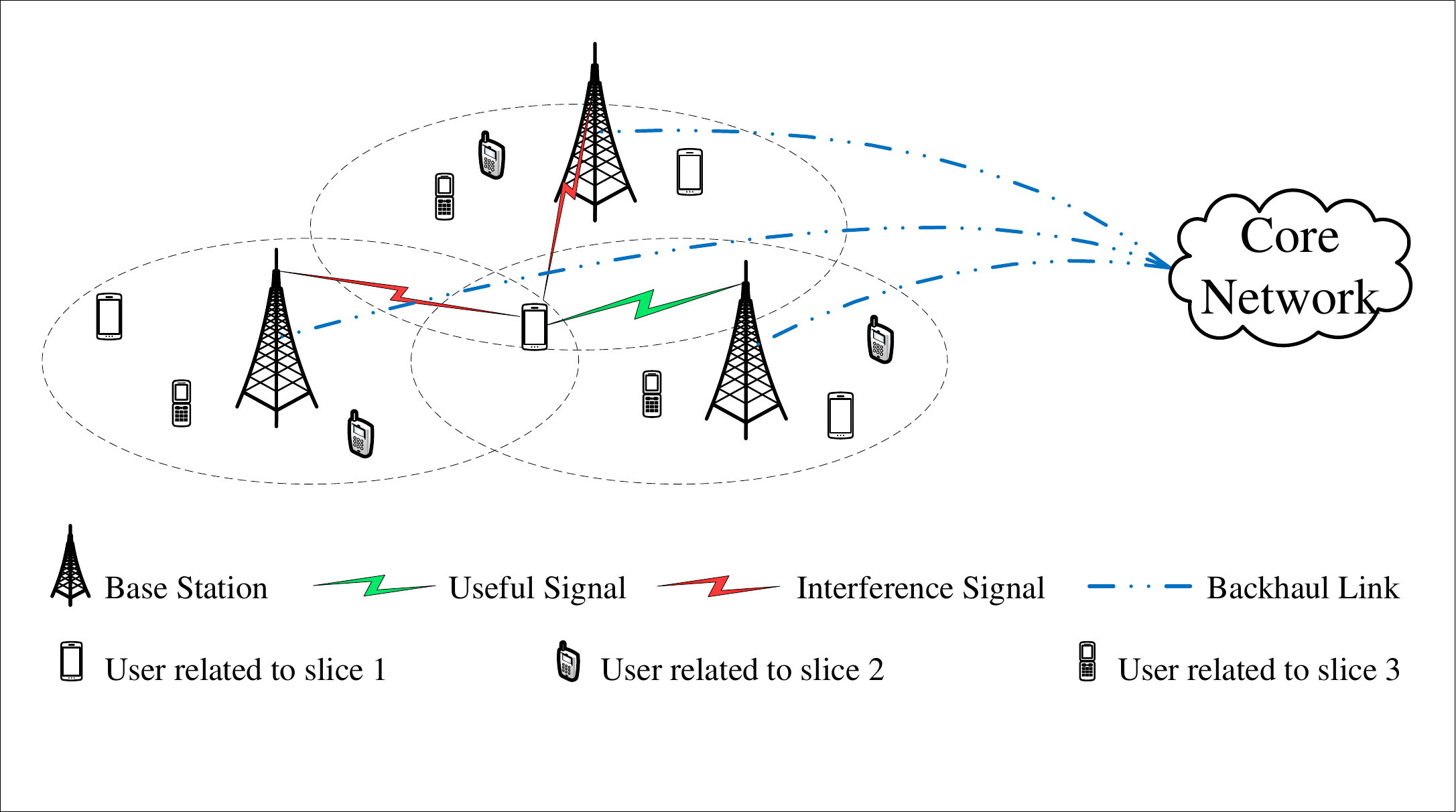}
	\caption{Illustration of the RAN domain in E2E NwS containing three slices in a coverage area BSs based on OFDMA.}
	\label{Fig RAN}
\end{figure}

The system architecture of the RAN domain is shown in Fig. \ref{Fig RAN}. We consider the downlink transmission of an orthogonal frequency division multiple access (OFDMA)-based multi-cell cellular network. The set of users is denoted by $\mathcal C=\left\{1,2,\dots,c,\dots,C\right\}$, where $C$ is the total number of users, and each user sends a slice request to the InP. The set of all BSs indicated by $\mathcal I=\left\{1,2,\dots,i,\dots,I\right\}$, where in this set $I$ is the total number of BS. Total bandwidth $B$ is divided into a set of the same subchannels $\mathcal{K}=\{1,2,\dots,k,\dots,K\}$, in which that $K$ is the total number of subchannels. Therefore, $B_k$ is the bandwidth of subchannel $k$. According to 3GPP, there are three main slice types in the 5G network: eMBB, uRLLC, and mMTC. Therefore, the set of slices is denoted by $\mathcal{S}=\left\{1,2,\dots,s,\dots,S\right\}$, where $S$ represents the total number of slices. Slice $s$ has a set of users $\mathcal{M}_{s}=\left\{1,2,\dots,m_{s},\dots,M_{s}\right\}$, where $m_s$ is the $m$-th user in slice $s$ and $M_s$ is the total number of users in slice $s$, therefore we have $\sum_{s\in\mathcal S}M_s=C$. We define a binary indicator $\delta^{i}_{m_s}$ as follows to determine the slice request by user $m_s$ from the InP:
\begin{align}
	\delta^{i}_{m_s}=\begin{cases}
		1, & \text{If the user $m_s$ at BS $i$ requests slice $s$,} 
		\\
		&\text{from InP;}
		\\
		0, & \text{Otherwise.}
	\end{cases}
\end{align}	
%
We assume that each user requests only one slice \cite{nouruzi2021online}. To this end, we consider the following constraint to ensure that each user can only request one slice from InP:
\begin{align}\label{C1}
	\text{C1}: \sum_{s\in\mathcal{S}}\delta^{i}_{m_s}=1, \forall i\in\mathcal{I}, \forall m_s\in\mathcal{M}_s.
\end{align}
Moreover, we define a binary variable $\xi^{i,k}_{m_s}$ for assigning subchannel $k$ by BS $i$ to user $m_s$ as follows:
\begin{align}
	\xi^{i,k}_{m_s}=\begin{cases}
		1, & \text{If BS $i$ allocates subchannel $k$,} 
		\\
		&\text{to user $m_s$;}
		\\
		0, & \text{Otherwise.}
	\end{cases}
\end{align}	
To ensure that each subchannel is allocated to just one user in each BS, we consider the following constraint:
\begin{align}\label{C2}
	\text{C2}: \sum_{s\in\mathcal S}\sum_{m_s\in\mathcal {M} _s}\xi^{i,k}_{m_s}\leqslant1, \forall i\in\mathcal{I}, \forall k\in\mathcal{K}.
\end{align}  
Let $p^{i,k}_{m_s}$ and $h^{i,k}_{m_s}$ denote the transmit power and the channel gain between BS $i$ and user $m_{s}$ on subchannel $k$, respectively. Accordingly, the transmit data rate from BS $i$ to user $m_{s}$ on subchannel $k$,  is calculated as follows:
\begin{align}\label{Data rate}
	&R^{i,k}_{m_s}=\delta^{i}_{m_s}B_{k}\log_{2}\left( 1+\frac{\xi^{i,k}_{m_s}p^{i,k}_{m_s}|h^{i,k}_{m_s}|^2}{I^{i,k}_{m_s}+\sigma^{2}}\right), \forall i\in \mathcal I, 
	\\ \nonumber
	&\forall k\in\mathcal K, 
	\forall m_s\in\mathcal {M}_s,
\end{align}
\begin{table}[h!]
	\centering
	\caption {Main notations}
	\label{Main notations}
	\scalebox{0.7}{
		\begin{tabular}{|c|l|}
			\hline
					\rowcolor[HTML]{38FFF8} 
			\textbf{Notation}                                     & \multicolumn{1}{c|}{\textbf{Definition}}                                                                                                                                                      \\ \hline
			\multicolumn{2}{|c|}{\textbf{Input Parameters}}                                                                                                                                                                                                       \\ \hline
			$\mathcal C$/$C$/$c$                                  & Set/number/index of users                                                                                                                                                                     \\ \hline
			$\mathcal I$/$I$/$i$                                  & Set/number/index of BSs                                                                                                                                                                       \\ \hline
			$\mathcal K$/$K$/$k$                                  & Set/number/index of subchannels                                                                                                                                                               \\ \hline
			$\mathcal S$/$S$/$s$                                  & Set/number/index of slices types                                                                                                                                                              \\ \hline
			$\mathcal {M}_s$/$M_s$/$m_s$                          & Set/number/index of users related to slice $s$                                                                                                                                                \\ \hline
			$B_{k}$                                               & Bandwidth of subchannel $k$                                                                                                                                                                   \\ \hline
			$p^{i,k}_{m_s}$                                       & \begin{tabular}[c]{@{}l@{}}Transmitted power from BS $i$ to user $m_{s}$ on subchannel $k$ \end{tabular}                                                                                   \\ \hline
			$h^{i,k}_{m_s}$                                       & \begin{tabular}[c]{@{}l@{}}Channel gain between BS $i$ and user $m_{s}$ on subchannel $k$\end{tabular}                                                                     \\ \hline
			$I^{i,k}_{m_s}$                                       & \begin{tabular}[c]{@{}l@{}}Inter-cell interference on user $m_s$ at BS $i$ over subchannel $k$\end{tabular}                                                                                 \\ \hline
			$\sigma^{2}$                                  & \begin{tabular}[c]{@{}l@{}}Power of additive white Gaussian noise (AWGN)\end{tabular}                                                           \\ \hline
			$P^i_{\text{max}}$                                    & Maximum transmit power of the BS $i$                                                                                                                                                          \\ \hline
			$\tilde h^{i,k}_{m_s}$                                & Estimated channel gain                                                                                                                                                                        \\ \hline
			$\epsilon^{i,k}_{m_s}$                                &Error of estimation channel gain                                                                                                                                                          \\ \hline
			$\Gamma^{i,k}_{m_s}$                                  & Channel gain uncertainty bound                                                                                                                                                                \\ \hline
			$\tilde{R}^{i,k}_{m_s}$                               & \begin{tabular}[c]{@{}l@{}}Transmitted data rate from BS $i$ to user $m_{s}$ on subchannel $k$ under \\ CSI uncertainty\end{tabular}                                                          \\ \hline
			$R^s_{\text{min}}$                                    & Minimum required data rate of slice $s$                                                                                                                                                       \\ \hline
			$x^i_{m_s}$                                           & Physical distance between BS $i$ and user $m_s$ in meter                                                                                                                                  \\ \hline
			$\nu$                                                 & Speed of light in meter per second                                                                                                                                                        \\ \hline
			$w_s'$                                                & Packet size in bits                                                                                                                                                                       \\ \hline
			$\mathcal G=(\mathcal N,\mathcal L)$                  & Network graph in core domain                                                                                                                                                                                 \\ \hline
			$\mathcal N$/$N$/$n$                                  & Set/number/index of nodes                                                                                                                                                                 \\ \hline
			$\mathcal L$/$L$/$l$                                  & Set/number/index of links                                                                                                                                                                     \\ \hline
			$BW_{n,n'}$                                           & \begin{tabular}[c]{@{}l@{}}Bandwidth capacity between nodes $n$ and $n'$ in bits per second\end{tabular}                                                                                   \\ \hline
			$\mathcal V$/$V$/$v$                                  & Set/number/index of VMs                                                                                                                                                                       \\ \hline
			$\mathcal P_{b,b'}$                                   & Set of all physical paths between nodes $b$ and $b'$                                                                                                                                          \\ \hline
			$\mathcal {Z}_{b,b'}^{v,v'}$                          & Set of virtual paths between VM $v$ and $v'$ on nodes $b$ and $b'$                                                                                                                            \\ \hline
			$\mathcal F$                                          & Set of all VNFs types                                                                                                                                                                         \\ \hline
			$q_f^{m_s}$                                           & \begin{tabular}[c]{@{}l@{}}Corresponding processing requirement for VNF $f$ in CPU cycle per \\ bit for each user $m_s$ of slice $s$\end{tabular}                                              \\ \hline
			$d^{f,m_s}_{v,b}$                                     & Processing delay of VNF $f$ on node $b$ for user $m_s$ of slice $s$                                                                                                                       \\ \hline
			$\alpha^{n,n'}_{\text{Prop}}$                         & Propagation delay between nodes $n$ and $n'$                                                                                                                                              \\ \hline
			$\tau_{\text{max}}^s$                                 & Maximum tolerable delay                                                                                                                                                                       \\ \hline
			$\Psi^{s}_{\text{Rev}}$                             & Unit price of data rate                                                                                                                                                                   \\ \hline
			$\Psi^{m_s}_{\text{Cost.RAN}}$                        & Unit cost of transmitted power by BS $i$                                                                                                                                                  \\ \hline
			$\Psi^{b,m_s}_{\text{Cost.core}}$                     & Unit cost of the node                                                                                                                                                                     \\ \hline
			$\Psi^{n,n',m_s}_{\text{Cost.core}}$                  & Unit cost of the link                                                                                                                                                                     
			\\ \hline
			$\Theta_{1},\Theta_{2}$                               & Scaling factors
			\\ \hline                                                                                                             
			\multicolumn{2}{|c|}{\textbf{Indicators/Variables/Matrix}}                                                               \\ \hline
			$\delta^{i}_{m_s}\in\{0,1\}$                          & \begin{tabular}[c]{@{}l@{}}Slice request indicator, that if user $m_s$ at BS $i$ requests slice $s$ from \\ InP it is 1, and otherwise 0\end{tabular}                                         \\ \hline
			$\xi^{i,k}_{m_s}\in\{0,1\}$                           & \begin{tabular}[c]{@{}l@{}}Subchannel allocation variable, that if BS $i$ is allocated subchannel $k$ \\ to user $m_s$ it is 1, and otherwise 0\end{tabular}                              \\ \hline
			$\mathbf{L}=[l_{n,n'}]\in\{0,1\}$                     & \begin{tabular}[c]{@{}l@{}}Connectivity matrix of the graph, that if nodes $n$ and $n'$ are connected \\ it is 1, and otherwise 0\end{tabular}                                                \\ \hline
			$\zeta^{n,n'}_{p_{b,b'}}\in\{0,1\}$                   & \begin{tabular}[c]{@{}l@{}}Binary indicator variable, that if the link between nodes $n$ and $n'$ is \\ in the path $p_{b,b'}$ it is 1, and otherwise 0\end{tabular}                          \\ \hline
			$\beta_{v,b}^{f^j,m_s}\in\{0,1\}$                     & \begin{tabular}[c]{@{}l@{}}Selection variable, that if VNF $f^j$ for user $m_s$ of slice $s$ is running \\ on VM $v$ in node $b$ it is 1, and otherwise 0\end{tabular}                        \\ \hline
			$\Upsilon^{m_s,z_{b,b'}^{v,v'}}_{p_{b,b'}}\in\{0,1\}$ & \begin{tabular}[c]{@{}l@{}}Decision variable, that if physical path $p_{b,b'}$ is chosen to transmit \\ the data for user $m_s$ from VM $v$ to $v'$ it is 1, and otherwise 0\end{tabular} \\ \hline
		\end{tabular}
	}
\end{table}
where $\sigma^{2}$ is the power of additive white Gaussian noise (AWGN), and $I^{i,k}_{m_s}$ is the inter-cell interference on user $m_s$ in BS $i$ over subchannel $k$, according to the following formula:
\begin{align}
	&I^{i,k}_{m_s}=\sum_{\substack{i'\in \mathcal{I}\\i'\neq i}}\sum_{\substack{m_s'\in\mathcal{M}_s\\m_s'\neq m_s}}\xi^{i',k}_{m_s'}p^{i',k}_{m_s'} |h^{i',k}_{m_s}|^2, \forall k\in\mathcal K.
\end{align}
To guarantee that the transmit power of each BS does not exceed its maximum transmit power $P^i_{\text{max}}$, the following constraint is proposed:
\begin{align}\label{C3}
	\text{C3}: \sum_{k\in \mathcal K}\sum_{s\in\mathcal S}\sum_{m_s\in\mathcal {M}_s}\delta^{i}_{m_s}\xi^{i,k}_{m_s}p^{i,k}_{m_s}\leqslant P^i_{\text{max}}, \forall i\in\mathcal{I}.
\end{align}
\subsection{Worst-case CSI Uncertainty Model}
In \eqref{Data rate}, perfect CSI is considered. In wireless communication, perfect CSI is not a valid assumption. In the BS, CSI uncertainty can occur by various factors such as mobility of users, estimation errors, hardware deficiencies, and delay in the feedback channel \cite{jumba2015energy,bjornson2012robust}. Perfect CSI at the BS is hard to obtain. Thus, to address this issue, we consider the CSI uncertainty at the BS. The imperfect CSI is given as follows \cite{zakeri2021robust}:
\begin{align}
	&h^{i,k}_{m_s}=\tilde h^{i,k}_{m_s}+\epsilon^{i,k}_{m_s}, \forall i\in \mathcal{I}, \forall k\in \mathcal{K}, \forall m_s\in \mathcal{M}_s,
\end{align}

where $\tilde h^{i,k}_{m_s}$ and $\epsilon^{i,k}_{m_s}$ are the estimated channel gain and error of estimation channel gain, respectively. The error of estimated channel gain are trapped in the bounded region. Thus, we have $h^{i,k}_{m_s}\in H^{i,k}_{m_s}$, where $H^{i,k}_{m_s}$ is expressed as follows:
\begin{align}
	&H^{i,k}_{m_s}\triangleq\left\{\tilde h^{i,k}_{m_s}+\epsilon^{i,k}_{m_s}\Big{|}\left|\epsilon^{i,k}_{m_s}\right|\leqslant\Gamma^{i,k}_{m_s}\right\}, \forall i\in\mathcal I, \forall k\in\mathcal K, 
	\\ \nonumber
	& m_s\in\mathcal {M}_s,
\end{align}	
where $\Gamma^{i,k}_{m_s}$ is the channel uncertainty bound, assumed to be a small constant. Therefore, the worst-case data rate of user $m_{s}$ under the CSI uncertainty can be formulated by \cite{zakeri2021robust}:
\begin{align}
	&\tilde{R}^{i}_{m_s}=\sum_{\substack{k'\in\mathcal K}}\min_{\left\{h^{i,k}_{m_s}\in H^{i,k}_{m_s} \right\}}R^{i,k'}_{m_s}, \forall i \in\mathcal{I}, \forall m_s\in\mathcal {M}_s.
\end{align}
Eventually, the total data rate of slice $s$ can be expressed as follows:
\begin{align}
	&\tilde{R}^s=\sum_{i\in\mathcal I}\sum_{m_s\in\mathcal{M}_s}\tilde{R}^{i}_{m_s}, \forall s\in\mathcal S.
\end{align}
Each slice $s\in\mathcal S$ requires a minimum data rate $R^s_{\text{min}}$. Therefore, to guarantee users’ QoS, we consider the following constraint:
\begin{align}\label{C4}
	\text{C4}: \sum_{i\in\mathcal I}\sum_{k\in\mathcal K}\sum_{m_s\in\mathcal {M}_s}\delta^{i}_{m_s}\xi^{i,k}_{m_s}\tilde{R}^s\geqslant R^s_{\text{min}}, \forall s\in\mathcal S.
\end{align}
\subsection{Delay Model in RAN}\label{RAN Delay}
We consider two types of delay in the RAN domain: propagation and transmission delay, which are computed in the following.
\subsubsection{Propagation Delay} 
We define $x^i_{m_s}$ as the physical distance between BS $i$ and user $m_s$ in meters and $\nu$ as the speed of light in meter per second. Therefore, the propagation delay between BS $i$  and user $m_s$ is given by:
\begin{align}
	D_{\text{Prop.RAN}}^{m_s}=\frac{x^i_{m_s}}{\nu}\delta^{i}_{m_s} \max_{k}\left\{\xi^{i,k}_{m_s}\right\}, \forall i\in\mathcal I, \forall m_s\in\mathcal {M}_s.
\end{align}
\subsubsection{Transmission Delay} 
Let $w'_{m_s}$ be the packet size in bits that is equal to the number of bits transmitted in one second. Accordingly, by transmitting data of user $m_s$ on subchannel $k$ by BS $i$, the transmission delay is calculated as follows:
\begin{align}
	D_{\text{Tran.RAN}}^{m_s}=\begin{cases}
		0, & \tilde{R}^{i}_{m_s}=0; 
		\\
		\frac{w'_{m_s}}{\tilde{R}^{i}_{m_s}}\delta^{i}_{m_s},&{\tilde{R}^{i}_{m_s}> 0}, \forall i\in\mathcal I, \forall m_s\in\mathcal {M}_s. 
	\end{cases}
\end{align}
\textbf{Remark 1.} \textit{It is worth noting that queue delay is often considered in problems involving mobile edge computing (MEC). Since we do not have the MEC in this paper, we do not model the queue delay in RAN and core domains in our system model similar to the existing works \cite{tang2019service,liu2020provisioning,mei20205g}.}
\subsection{Core Network Explanation} \label{Core}	
\begin{figure}[h!]
	\centering
	\includegraphics[width=0.94\linewidth]{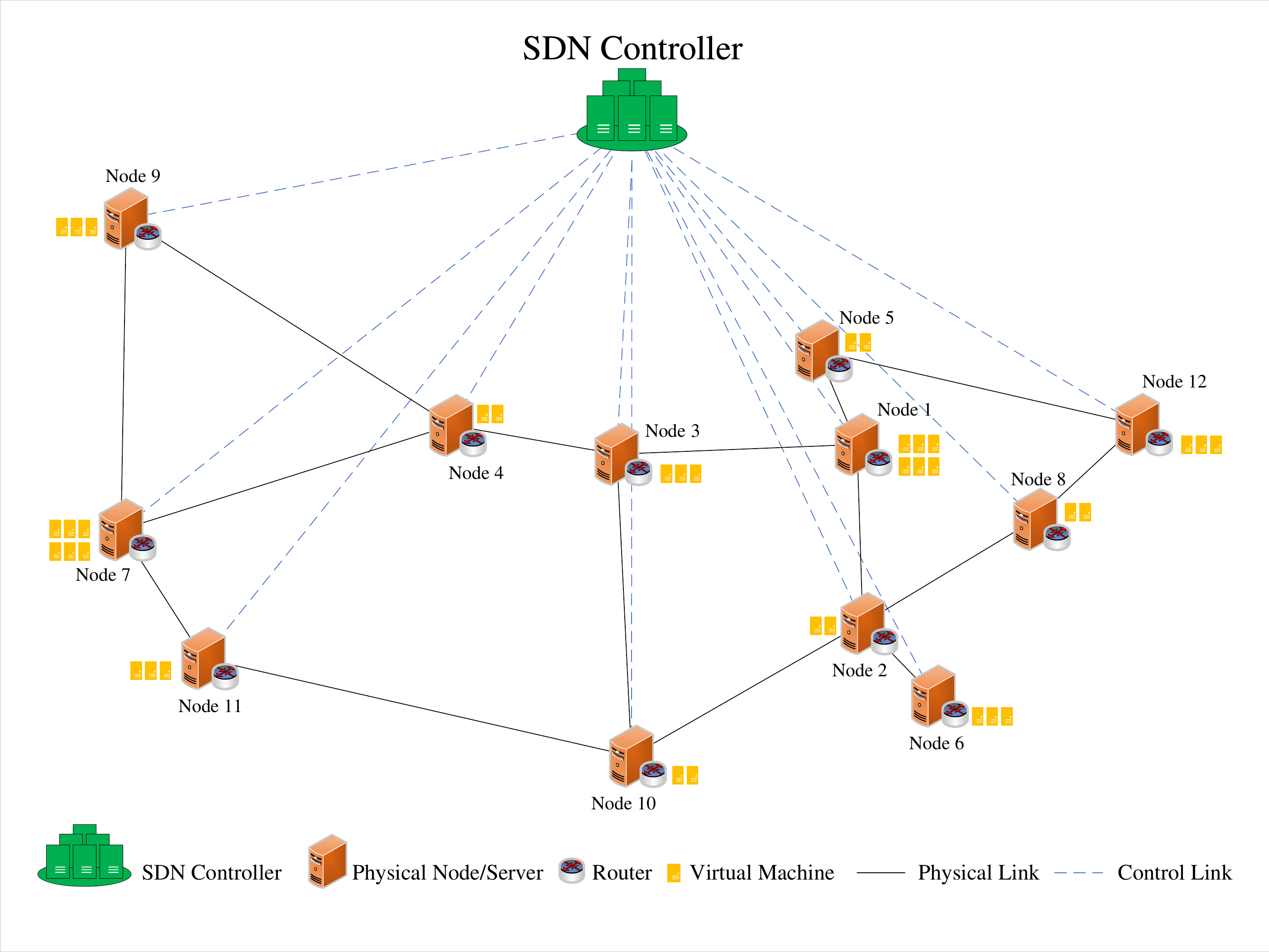}
	\caption{Illustration of the core domain in E2E NwS based on \textit{Abilene} network topology with 12 nodes and 15 links from \textit{SNDlib} \cite{sndlib}.}
	\label{Fig Core}
\end{figure}

The system architecture of the core domain is shown in Fig. \ref{Fig Core}. We consider the core network as a graph $\mathcal G=(\mathcal N,\mathcal L)$, where $\mathcal N=\left\{1,2,\dots,n,\dots,N\right\}$ represents the set of nodes, and $\mathcal L=\left\{1,2,\dots,l,\dots,L\right\}$ represents the set of links. Moreover, the total number of nodes and links in these sets are denoted by $N$ and $L$, respectively. In addition, let $\mathbf{L}=[l_{n,n'}]$ be the connectivity matrix of physical links, that is defined as:
\begin{align}
	l_{n,n'}=\begin{cases}
		1, & \text{If nodes $n$ and $n'$ are connected;} 
		\\
		0, & \text{Otherwise.}
	\end{cases}
\end{align}	
We denote the total resource capacity of each node $n\in\mathcal N$ as $\boldsymbol{r^n}=[r^n_{\text{CPU}},r^n_{\text{RAM}},r^n_{\text{Stor}}]$, where $r^n_{\text{CPU}}$, $r^n_{\text{RAM}}$, and $r^n_{\text{Stor}}$ indicate the CPU, RAM, and storage capacities of node $n$, respectively. Moreover, we denote the limited bandwidth capacity between nodes $n$ and $n'$ in bits per second by $BW_{n,n'}$. In addition, we consider that each node $b$ hosts several VMs which are denoted by $\mathcal V_{b}=\{1_b,\dots,v_b,\dots,V_b\}$, where $V_b$ is the total number of VMs in node $b$. Accordingly the set of total VMs in the network is denoted by $\mathcal V_{\text{Total}}=\cup_{b=1}^N\mathcal V_b$. The maximum number of VMs on each node is indicated by $V_{\text{max}}$. Moreover, each VM $v$ on node $b$ has specific CPU, RAM, and storage resources, that are represented by $\boldsymbol{r^{v,b}}=[r^{v,b}_{\text{CPU}},r^{v,b}_{\text{RAM}},r^{v,b}_{\text{Stor}}]$. Let $\mathcal P_{b,b'}=\{1_{b,b'},\dots,p_{b,b'},\dots,P_{b,b'}\}$ be the set of physical paths between nodes $b$ and $b'$, where $p_{b,b'}$ and $P_{b,b'}$ denote the $p$-th path and total paths between nodes $b$ and $b'$, respectively. We define a binary indicator $\zeta^{n,n'}_{p_{b,b'}}$ as follows to determine which of the physical links are in a path:
\begin{align}
	\zeta^{n,n'}_{p_{b,b'}}=\begin{cases}
		1, & \text{If the link between nodes $n$ and $n'$,} 
		\\
		&\text{is in path $p_{b,b'}$;}
		\\
		0, & \text{Otherwise.}
	\end{cases}
\end{align}	
Furthermore, the set of virtual paths between VMs $v$ and $v'$ on nodes $b$ and $b'$ is represented by $\mathcal {Z}_{b,b'}^{v,v'}=\{1_{b,b'}^{v,v'},\dots,z_{b,b'}^{v,v'},\dots,Z_{b,b'}^{v,v'}\}$, where $z_{b,b'}^{v,v'}$  is the $z$-th path in this set \cite{ebrahimi2020joint}, \cite{tajiki2018joint}. Each slice consists of several services such as web service, voice over internet protocol (VoIP), video streaming, cloud gaming, etc. Different VNFs must implement these various services \cite{savi2019impact}. We consider VNFs as the network address translator (NAT), firewall (FW), traffic monitor (TM), wide area network (WAN) optimization controller (WOC), intrusion detection prevention system (IDPS), and video optimization controller (VOC). All VNFs types are denoted

by $\mathcal F=\{1,2,\dots,f,\dots,F\}$, Where $F$ is the total VNF types. For each slice $s\in\mathcal S$, several VNFs must be run in VMs. Thus, the VNF chain of slice $s$ is represented by $\mathcal {F}_s=\{f^1_s,f^2_s,\dots,f^j_s,\dots,F_s^{\text{Total}}\}$, where $F_s^{\text{Total}}$ is the total number of VNFs required for slice $s$. The order of running of VNFs for each slice $s\in\mathcal{S}$ is already known. Let $\beta_{v,b}^{f^j,m_s}$ be the binary decision variable to determine that the $j$-th function for user $m_s$ of slice $s$ is running on VM $v$ in node $b$, that we define it as follows:
\begin{align}
	\beta_{v,b}^{f^j,m_s}=\begin{cases}
		1, & \text{If VNF $f^j$ for user $m_s$ of slice $s$,} 
		\\
		&\text{is running on VM $v$ in node $b$;}
		\\
		0, & \text{Otherwise.}
	\end{cases}
\end{align}
We consider the following constraint to ensure that each VNF completely runs or is assigned to one VM: 
\begin{align}\label{C5}
	\text{C5}: \sum_{v\in\mathcal V}\sum_{b\in\mathcal N}\beta_{v,b}^{f^j,m_s}=1, \forall f^j\in\mathcal {F}_s, \forall m_s\in\mathcal {M}_s.
\end{align}
To map the virtual path $z_{b,b'}^{v,v'}$ to the physical path $p_{b,b'}$, we define a binary decision variable $\Upsilon^{m_s,z_{b,b'}^{v,v'}}_{p_{b,b'}}$ as below:
\begin{align}
	\Upsilon^{m_s,z_{b,b'}^{v,v'}}_{p_{b,b'}}=\begin{cases}
		1, & \text{If the physical path $p_{b,b'}$ is chosen to,}
		\\
		&\text{transmit the data rate for user $m_s$,}
		\\
		&\text{of slice $s$ from VM $v$ to $v'$;}
		\\
		0, & \text{Otherwise.}
	\end{cases}
\end{align}
For each virtual path just one physical path is selected. Accordingly, we define the following constraint:
\begin{align}\label{C5}
	\text{C6}: \sum_{p_{b,b'}\in\mathcal{P}_{b,b'}}\Upsilon^{m_s,z_{b,b'}^{v,v'}}_{p_{b,b'}}=1, \forall m_s\in\mathcal {M}_s, \forall z_{b,b'}^{v,v'}\in\mathcal {Z}_{b,b'}^{v,v'}.
\end{align}
In addition, because each  VNF needs corresponding computation resources, let  $q_f^{m_s}$ denote the corresponding processing requirement for VNF $f$ in CPU cycle per bit for each user $m_s$ of slice $s$. Hence, we have a set of the related processing requirements as below:
\begin{align}
Q_{m_s}=\left\{q_f^{m_s}\right\}, \forall m_s\in\mathcal {M}_s, \forall f\in\mathcal {F}_s.	
\end{align}
\subsection{Delay Model in Core Network}\label{Core Delay}
In the following, we compute three types of delay on the core domain: Processing, transmission, and propagation delay \cite{nouruzi2021online}.
\subsubsection{Processing Delay}
We define $d^{f,m_s}_{v,b}$ as the processing delay of VNF $f$ on node $b$ for user $m_s$ of slice $s$ in VM $v$ that is defined as follows:
\begin{align}
	d^{f,m_s}_{v,b}=\frac{w'_{m_s}q_f^{m_s}}{r^{v,b}_{\text{CPU}}}, \forall f\in\mathcal {F}_s, \forall m_s\in \mathcal {M}_s, \forall v\in\mathcal{V}_b, \forall b\in\mathcal{N}.
\end{align}
Therefore, the processing delay for user $m_s$ can be expressed as follows:
\begin{align}
	&D_{\text{Proc.core}}^{m_s}=\sum_{j=1}^{F_s}\sum_{v\in\mathcal V}\sum_{b\in N}d^{f^j,m_s}_{v,b}\beta_{v,b}^{f^j,m_s}, \forall m_s\in\mathcal {M}_s.
\end{align}
\subsubsection{Propagation Delay}
Let $\alpha^{n,n'}_{\text{Prop}}$ be the propagation delay between nodes $n$ and $n'$. Therefore, we can write: 
\begin{align}
	\alpha^{n,n'}_{\text{Prop}}=\frac{x_{n,n'}}{\nu}, \forall n,n'\in\mathcal{N},	
\end{align}
where $x_{n,n'}$ is the distance of the physical link between node $n$ and $n'$. As a result, the total propagation delay for user $m_s$ of slice $s$ is obtained by:
\begin{align}
&D_{\text{Prop.core}}^{m_s}=\sum_{j=1}^{F_s^{\text{Total}}-1}\sum_{\substack{n,n',b,b'\in N\\v,v'\in\mathcal {V}_b\\p_{b,b'}\in\mathcal {P}_{b,b'}}}\alpha^{n,n'}_{\text{Prop}}\zeta^{n,n'}_{p_{b,b'}}\Upsilon^{m_s,z_{b,b'}^{v,v'}}_{p_{b,b'}}\beta_{v,b}^{f^j,m_s}
\\ \nonumber
&\beta_{v',b'}^{f^{j+1},m_s}, \forall m_s\in\mathcal{M}_s.		
\end{align}
\subsubsection{Transmission Delay} 
The transmission delay for user $m_s$ of slice $s$ is formulated by:
\begin{align}
	&D_{\text{Tran.core}}^{m_s}=\sum_{j=1}^{F_s}\sum_{\substack{n,n',b,b'\in N\\v,v'\in\mathcal{V}_b\\p_{b,b'}\in\mathcal {P}_{b,b'}}}\frac{\tilde{w}_{m_s}}{BW_{n,n'}}\zeta^{n,n'}_{p_{b,b'}}\Upsilon^{m_s,z_{b,b'}^{v,v'}}_{p_{b,b'}}\beta_{v,b}^{f^j,m_s} 
	\\ \nonumber
	&\beta_{v',b'}^{f^{j+1},m_s}, \forall m_s\in\mathcal{M}_s,
\end{align}	
where $\tilde{w}_{m_s}$ is the uncertain data rate in bits per second.
\subsection{E2E Delay Model}
Based on Subsections \ref{RAN Delay} and \ref{Core Delay}, the total E2E delay for each user $m_s$ is obtained as follows:
\begin{align}
	&D_{\text{Total}}^{m_s}=D_{\text{Proc.core}}^{m_s}+D_{\text{Prop.core}}^{m_s}+D_{\text{Tran.core}}^{m_s}+
	\\ \nonumber
	&D_{\text{Prop.RAN}}^{m_s}+D_{\text{Tran.RAN}}^{m_s}, \forall m_s\in\mathcal{M}_s.
\end{align}
To make sure that the total E2E delay for each user $m_s$ is less than the maximum tolerable delay time $\tau_{\text{max}}^s$, we use the following constraint:
\begin{align}\label{C7}
	\text{C7}: D_{\text{Total}}^{m_s}\leqslant\tau_{\text{max}}^s, \forall m_s\in\mathcal {M}_s, \forall s\in\mathcal S.
\end{align}
\subsection{Demand Uncertainty Model}
In the proposed system model, we assume that the number of users requests for each slice can be random. In addition, the data rate requests in each slice can be stochastic too. For this purpose, the data rate requested of virtual links (VLs) by user $m_s$ in each slice is denoted by $\tilde{w}_{m_s}$ that it is a random variable with a uniform distribution and considered as $\tilde {w}_{m_s}\in[\bar{w}_{m_s}-\hat{w}_{m_s},\bar{w}_{m_s}+\hat{w}_{m_s}]$, where $\bar{w}_{m_s}$ refers to the nominal bandwidth demand values or the mean of $\tilde{w}_{m_s}$, that is denoted by $\mathbb{E}[\tilde{w}_{m_s}]=\bar{w}_{m_s}$, and $\hat{w}_{m_s}\geqslant0$ is the maximum bandwidth demand deviation or the standard deviation of $\tilde{w}_s$, that is indicated by $\text{Var}[\tilde{w}_{m_s}]={\hat{w}^2_{m_s}}$. Note that there is no information about the exact value of $\tilde{w}_{m_s}$; we only know its mean and variance values \cite{hosseini2019probabilistic}. We can model the demand uncertainty as follows \cite{baumgartner2017network,gholamipour2021online}:
\begin{align}\label{Demand}
	&\sum_{m_s\in\mathcal {M}_s}\sum_{n\in\mathcal N}\sum_{n'\in\mathcal N}\sum_{b\in\mathcal P_{b,b'} }\delta^{i}_{m_s}\zeta^{n,n'}_{p_{b,b'}}\Upsilon^{m_s,z_{b,b'}^{v,v'}}_{p_{b,b'}}\tilde{w}_{m_s}\leqslant BW_{n,n'},   
	\\ \nonumber
	&\forall i\in\mathcal I, \forall p_{b,b'}\in\mathcal {P}_{b,b'}, z_{b,b'}^{v,v'}\in\mathcal {Z}_{b,b'}^{v,v'}.
\end{align}
Note that \eqref{Demand} has a stochastic variable where we can not solve directly. To this end, the stochastic variable is changed into the deterministic variable. Therefore, we have the following constraint \cite{gholamipour2021online}:
\begin{align}\label{C8}
&\text{C8}: \sum_{m_s\in\mathcal {M}_s}\sum_{n\in\mathcal N}\sum_{n'\in\mathcal N}\sum_{b\in\mathcal P_{b,b'} }\delta^{i}_{m_s}\zeta^{n,n'}_{p_{b,b'}}\Upsilon^{m_s,z_{b,b'}^{v,v'}}_{p_{b,b'}}\bar{w}_{m_s}+  
\\ \nonumber 
&\max_{|\kappa_{m_s}|\leqslant\tilde{w}_{m_s}}\Bigg\{\sum_{m_s\in\mathcal {M}_s}\sum_{n\in\mathcal N}\sum_{n'\in\mathcal N}\sum_{b\in\mathcal P_{b,b'} }\delta^{i}_{m_s}\zeta^{n,n'}_{p_{b,b'}}\Upsilon^{m_s,z_{b,b'}^{v,v'}}_{p_{b,b'}} 
\\ \nonumber
&\kappa_{m_s}\hat{w}_{m_s}\Bigg\}\leqslant BW_{n,n'}, \forall i\in\mathcal I, \forall p_{b,b'}\in\mathcal {P}_{b,b'}, z_{b,b'}^{v,v'}\in\mathcal {Z}_{b,b'}^{v,v'},
\end{align}
where $\kappa_{m_s}$ is a auxiliary variable. In reality, constraint $\text{C8}$ is for the mean of the requested data rate plus the maximum standard deviation of the requested data rate to be less than the available capacity.

\section{Problem Formulation}\label{IV}
We aim to maximize the utility of the InP where each slice request is handled by the InP. In the following subsections, we formulate the revenue, cost, and utility functions, respectively.
\subsection{Revenue Function Model}
Let $\Psi^{s}_{\text{Rev}}$ represent the unit price of the data rate for slice $s$, that has dimension \$/Mbps. If the InP accepts the user's slice request and allocates the data rate, it can generate revenue. To this end, the revenue function $\mathbb{U}^{s}_{\text{Rev}}$ is obtained by:
\begin{align}
	&\mathbb{U}^{s}_{\text{Rev}}=\Psi^{s}_{\text{Rev}}\tilde{R}^s, \forall s\in\mathcal{S}.
\end{align}
\subsection{Cost Function Model}
In this paper, we consider two types of cost for RAN and core domains in our system model, that in the following parts the detail of them is introduced.
\subsubsection{Cost Function in RAN}
Let $\Psi^{m_s}_{\text{Cost.RAN}}$ be the unit cost of the transmitted power by BS on each subchannel for user $m_s$ of slice $s$ with dimension \$/Watt/Hz. We denote the cost function in RAN domain by $\mathbb{U}^{s}_{\text{Cost.RAN}}$, and it can be formulated as follows:
\begin{align}\label{Cost.RAN}
	&\mathbb{U}^{s}_{\text{Cost.RAN}}=\sum_{m_s\in\mathcal {M}_s}\sum_{k\in\mathcal K}\xi^{i,k}_{m_s}p^{i,k}_{m_s}\Psi^{m_s}_{\text{Cost.RAN}}, \forall i\in\mathcal I.
\end{align}
\subsubsection{Cost Function in Core Network}
Let $\Psi^{b,m_s}_{\text{Cost.core}}$ and $\Psi^{n,n',m_s}_{\text{Cost.core}}$ be the unit cost of the node and link, respectively. Accordingly, we define $\mathbb{U}^{s}_{\text{Cost.core}}$ as the cost of utilization of network resources and includes the node and link bandwidth in the core domain. Thus, it can be expressed as follows:
\begin{align}\label{Cost.core}
	&\mathbb{U}^{s}_{\text{Cost.core}}=\sum_{\substack{m_s\in\mathcal {M}_s\\b\in\mathcal N\\v\in\mathcal V\\f\in\mathcal{F}_s}}w'_{m_s}\beta^{f,m_s}_{v,b}q_f^{m_s}\Psi^{b,m_s}_{\text{Cost.core}}+ 
	\\ \nonumber
	&\sum_{\substack{m_s\in\mathcal {M}_s\\n,n',b,b'\in\mathcal N\\p_{b,b'}\in \mathcal P_{b,b'}}}w'_{m_s}\Upsilon^{s,z_{b,b'}^{v,v'}}_{p_{b,b'}}\beta_{v,b}^{f,m_s}\beta_{v',b'}^{f',m_s}\zeta^{n,n'}_{p_{b,b'}}\Psi^{n,n',m_s}_{\text{Cost.core}}.
\end{align}
\subsubsection{Total Cost Function} 
Finally, based on \eqref{Cost.RAN} and \eqref{Cost.core}, we can derive the total cost of slice $s$ in our network as follows:
\begin{align}
	\mathbb{U}^{s}_{\text{Cost}}=\mathbb{U}^{s}_{\text{Cost.RAN}}+\mathbb{U}^{s}_{\text{Cost.core}}, \forall s\in\mathcal{S}.
\end{align}
\subsection{Utility Function Model}
The utility function $\mathbb{U}^{s}$ is obtained from the difference between revenue and cost functions in the RAN and core domains. Therefore, it is computed as follows:
\begin{align}
	&\mathbb{U}^{s}=\Theta_{1}\mathbb{U}^{s}_{\text{Rev}}-\Theta_{2}\mathbb{U}^{s}_{\text{Cost}}, \forall s\in\mathcal{S}.
\end{align}
where $\Theta_{1},\Theta_{2}\geqslant 0$ are scaling factors and are used for balancing and scaling the revenue and costs of different resource types, respectively. At this point, based on the description in the previous sections, the  optimization problem in E2E NwS under demand and CSI uncertainties can be written as follows:
\begin{subequations}
	\begin{align}
&\max_{k,p,\beta,\Upsilon}\sum_{s\in S}\mathbb{U}^{s}\label{Optimization_Problem}
\\  \textbf{s.t.}~~&\text{C1}-\text{C8} \label{C1 until C8}
\\&\label{C9}\text{C9}: \xi^{i,k}_{m_s}\in\{0,1\}, \forall i\in\mathcal{I}, \forall k\in\mathcal{K}, \forall m_s\in\mathcal {M}_s.
\\&\label{C10}\text{C10}: \beta_{v,b}^{f^j,m_s}\in\{0,1\}, \forall f^j\in\mathcal F, \forall m_s\in\mathcal {M}_s,
\\ \nonumber
 &\forall v\in\mathcal V_{b}, \forall b\in\mathcal N.
\\&\label{C11}\text{C11}:\Upsilon^{s,z_{b,b'}^{v,v'}}_{p_{b,b'}}\in\{0,1\}, \forall s\in\mathcal S, \forall z_{b,b'}^{v,v'}\in\mathcal {Z}_{b,b'}^{v,v'}, 
\\ \nonumber
&\forall p_{b,b'}\in\mathcal  P_{b,b'},
\end{align}
\end{subequations}
where in Section \ref{III} the details of constraints $\text{C1-C8}$ are studied. In addition, constraints $\text{C9-C11}$ are used to ensure that decision variables are binary. The proposed problem formulation \eqref{Optimization_Problem}-\eqref{C11} is a difficult non-convex mixed-integer non-linear programming.
\section{Solution}\label{V}
Because of the complexity of the problem presented in Section \ref{IV}, we apply a robust method to solve our problem. In problems that include information uncertainty, mainly heuristic methods, $\Gamma$-robustness, and ML algorithms are used. In recent years, the use of ML algorithms to meet various challenges in mobile networks to improve performance and compatibility has increased significantly. In addition, many recent studies show that ML-based resource allocation is more effective than conventional methods \cite{franccois2018introduction,liang2019deep}. To this end, we employ several DRL algorithms \cite{dong2020deep} to solve our resource allocation problem. Model-free DRL algorithms such as DQN and double DQN are value-based methods where the Q-values are estimated with lower variance. These methods can not be used for problems with continuous action spaces. The best way to deal with this challenge is to apply policy gradient-based RL methods that can handle problems with continuous action space by learning deterministic/stochastic policies. The goal of these techniques is to optimize a policy based on the gradient of the expected reward. Nevertheless, these methods have very slow convergence. The deterministic policy gradient (DPG) algorithm \cite{silver2014deterministic} uses a learned approximation of the action-value (Q) function to approximate action-value gradients. The deep DPG (DDPG) \cite{lillicrap2015continuous} is an off-policy algorithm that uses the actor-critic method \cite{konda2000actor} to manage the continuous action spaces and to find the solution; this method needs a large number of training episodes, like in model-free RL algorithms. Based on DPG, DDPG employs a parameterized actor function to deterministically map states to specific actions while keeping DQN learning as the critic \cite{nguyen2020deep}. To address the slow convergence problem, the DDPG combines both features of policy-based and value-based algorithms to handle the continuous and large state/action spaces. \cite{heess2015memory} developed the DDPG approach to recurrent DPG (RDPG) by adding long short-term memory (LSTM)\footnote{LSTM is an artificial RNN structure that is used in deep learning. LSTM includes feedback connections, unlike standard feedforward neural networks.} \cite{hochreiter1997long} to solve the problems with continuous action spaces under partial observation. In partially observable systems the agent does not have the full state information (i.e., channel gains and bandwidth demand in our problem). In uncertain systems, the agent knows the information with bounded error. Due to uncertainties in the proposed problem formulation, we employ the RDPG algorithm as the main approach. RDPG exploits recurrent neural networks (RNNs) feed-forward networks. In other words, by using the RNN instead of the feed-forward approach, we are able to learn from history. Moreover, we consider the SAC\cite{haarnoja2018soft}, DDPG, distributed, and greedy \cite{agarwal2018joint} algorithms as baselines.
\subsection{RDPG Algorithm}
We consider a standard DRL setup in which the agent interacts with the environment $E$ in discrete time slots. Note that each time slot $t$ is equal to one second. The MDP \cite{bellman1957markovian} is a sequential decision-making process suitable for a fully observed and stochastic environment with additive rewards and the Markovian transition model. The typical RL problem is modeled as the MDP. In the fully observed MDP, when we have access to state $s_t$, the action-value function is expressed as the expected future discounted reward. Because the accurate state information is not available in the partially-observed MDP (POMDP) \cite{aastrom1965optimal}, it employs knowledge of actions and observations from previous time-steps to improve current observations. In this paper, we model our environment as the POMDP. The DRL algorithm includes an agent, a set of environment states $\mathcal{S}$, a set of actions $\mathcal{A}$, an initial state distribution $p_0(s_0)$, a transition function $p(s_{t+1}|s_t,a_t)$, and the reward function $r(s_t,a_t)$. In every time slot $t$ the agent receives an observation $o_t$, performs an action $a_t$ and gets a reward $r_t$. Because the agent is unable to observe state $s_t$ directly, it receives observations from the set $\mathcal{O}$ conditioned on the underlying state $p(o_t,s_t)$. In principle, the optimal agent may need access to the entire histories of observations represented by $h_t=(o_1,a_1,o_2,a_2,\dots,o_{t-1},a_{t-1},o_t)$. Therefore, the goal of the agent is to learn a policy $\pi(h_t)$ that maps from the history to the distribution of actions $P(\mathcal{A})$, and it maximizes the expected discounted reward. In the RDPG method, the policy is dependent on the whole history. The optimal policy and the associated action-value function are functions of the entire preceding observation-action history $h_t$. 
\\According to our optimization problem, we define the agent, the state space $\mathcal{S}$, the action space $\mathcal{A}$, and the reward function $r$ as follows:
\\$\bullet$ \textbf{Agent}: 
In our model, the SDN controller is assumed as the agent to select the actions form action spaces by considering network states. The agent receives a reward for each chosen action, and then the network’s state changes to the next state as time evolves. If the agent can not satisfy the constraints \eqref{C1 until C8}-\eqref{C11}, it will be punished with a negative reward. Also, if the agent does the good action, it will receive a positive reward.
\\$\bullet$ \textbf{System States}: 
Learning decisions are made based on the system state, which is an abstraction of the environment. The channel gain and bandwidth are the most significant parameters on the state of our system model environment. Thus, the system state $\mathcal{S}_t$ defined as the channel gain, link bandwidth, and uncertain data rate requested that can be expressed as:
\begin{align}
	\mathcal{S}_t=\left(h^{i,k}_{m_s},BW_{n,n'},\tilde {w}_{m_s}\right).
\end{align}
\\$\bullet$ \textbf{Action}: 
Since the network operates in a new state and transients from the current state of the network, the learner takes action based on its state. The action space $\mathcal{A}$ includes the transmit power from BS $i$ to user $m_{s}$ of slice $s$ on subchannel $k$, all the subchannels, paths, nodes, and VMs on the nodes. Therefore, the set of all actions can be expressed as: 
\begin{align}
	\mathcal{A}_t=\left(p^{i,k}_{m_s},\mathcal K,\mathcal P_{b,b'},\mathcal N,\mathcal V_{\text{Total}}\right).
\end{align}	
\begin{algorithm}[h!]
	\small
	\renewcommand{\arraystretch}{0.4}
	\caption{RDPG Algorithm \cite{heess2015memory}}
	\label{RDPG}
	\textbf{Input:} Initialize weights of actor and critic networks, $\mu^\theta(h_t)$ and $Q^\omega(h_t,a_t)$, with parameters $\theta$ and $\omega$. 
	\\\textbf{Input:} Initialize target network weights of actor and critic networks, $Q^{\omega'}$ and $\mu^{\theta'}$, with weights $\omega'\leftarrow\omega$, $\theta'\leftarrow\theta$
	\\\textbf{Input:} Initialize the replay buffer $\mathcal{B}$
	\\\For{episodes=1 to E}{
		Initialize empty history $h_0$
		\\\For{t=1 to $T$}{
			Receive observation $o_t$
			\\Add previous action and observation to history ($h_t\leftarrow h_{t-1}, a_{t-1}, o_t$)
			\\Based on the history at the time slot $t$ take action $a_t=\mu^{\theta}(h_t)+\epsilon$ ($\epsilon$: exploration noise)
		}
		Store the sequence $(o_1,a_1,r_1,\dots,o_T,a_T,r_T)$ in $\mathcal{B}$ 
		\\Sample a minibatch of $N$ episodes $(o_1^i,a_1^i,r_1^i,\dots,o_T^i,a_T^i,r_T^i)_{i=1,\dots,N}$ from $\mathcal{B}$
		\\Build histories $h_t^i=(o_1^i,a_1^i,\dots,a_{t-1}^i,o_t^i)$  
		\\Calculate the target values for each sample episode $(y_1^i,\dots,y_T^i)$ using the recurrent target networks:
		\begin{align}
			&\nonumber y_t^i=r_t^i+\gamma Q^{\omega'}\left(h_{t+1}^i,\mu^{\theta'}\left(h_{t+1}^i\right)\right)
		\end{align}
		\\Calculate critic update using backpropagation through time (BPTT):
		\begin{align}
			&\nonumber \Delta\omega=\frac{1}{NT}\sum_i\sum_t\left(y_t^i-Q^\omega\left(h_t^i,a_t^i\right)\right)\frac{\partial Q^\omega(h_t^i,a_t^i)}{\partial \omega}
		\end{align}
		\\Calculate actor update by using BPTT:
		\begin{align}
			&\nonumber \Delta\theta=\frac{1}{NT}\sum_i\sum_t\frac{\partial Q^\omega\left(h_t^i,\mu^\theta\left(h_t^i\right)\right)}{\partial a}\frac{\partial \mu^\theta\left(h_t^i\right)}{\partial \theta}
		\end{align}
		\\Update actor and critic using Adam optimizer \cite{kingma2014adam}
		\\Update the actor and critic target networks with the period $\tau$:
		\begin{align}
			&\nonumber\omega'\leftarrow\tau\omega+(1-\tau)\omega'\\
			&\nonumber\theta'\leftarrow\tau\theta+(1-\tau)\theta'
		\end{align}
	}
\end{algorithm}
\\$\bullet$ \textbf{Reward Function}: 
The agent receives the reward after taking action, which further reward improves network performance. In our optimization problem, the goal is to maximize the utility of the InP. Hence, the reward function is denoted as follows:
\begin{align}
	r_{t}\left(s_t,a_t\right)=\mathfrak{u}\mathbb{U}^s,
\end{align}
where $\mathfrak{u}$ is a coefficient factor. Moreover, the state of the system will change based on the actions that users take according to the system state. For example, positive rewards will be awarded if the agent takes a good action; otherwise, negative rewards will be awarded.
\\The following formula is considered to maximize the discounted expected cumulative reward:
\begin{align}
	J=\mathbb{E}_{\tau}\left[\sum_{t=1}^\infty\gamma^{t-1}r\left(s_t,a_t\right)\right],
\end{align}
where $\gamma\in[0,1]$ is a discount factor, and $\tau$ is set of a trajectories with length $L$ denoted by:
\begin{align}
	\tau=(s_1,o_1,a_1,s_2,o_2,a_2,\dots,s_L,o_L,a_L),
\end{align}
where the trajectory $\tau$ is computed from the trajectory distribution influenced by policy $\pi$ as follows:
\begin{align}
	\pi:p(s_1)\prod_{i=1}^{L}p(o_i|s_i)\pi(a_i|h_i)p(s_{i+1}|s_i,a_i).
\end{align}
In the employed algorithm, we use the action-value function $Q^\pi$. Therefore, $Q^\pi$ in terms of $h$ is expressed as follows:
\begin{align}
&Q^\pi\left(h_t,a_t\right)=\mathbb{E}_{s_t|h_t}\left[r_t\left(s_t,a_t\right)\right]+
\\ \nonumber &\mathbb{E}_{\tau>t|h_t,a_t}\left[\sum_{i=1}^L\gamma^ir\left(s_{t+i},a_{t+i}\right)\right],
\end{align}
where $\tau>t=(s_{t+1},o_{t+1},a_{t+1},\dots)$ is the future trajectory after $t$. The policy is updated as below:
\begin{align}
	&\frac{\partial J\left(\theta\right)}{\partial\theta}=\mathbb{E}_\tau\left[\sum_{t=1}^{\infty}\gamma^{t-1}\frac{\partial Q^\omega\left(h_t,a\right)}{\partial a}\Bigg{|}_{a=\mu^\theta\left(h_t\right)}\frac{\partial\mu^\theta\left(h_t\right)}{\partial\theta}\right],
\end{align}
where $Q^\omega$ is a recurrent network with parameters $\omega$.
Algorithm \ref{RDPG} is proposed to better understand the concept of the RDPG method. 
\subsection{DDPG Algorithm}
The DDPG method is a model-free, and off-policy-based RL approaches where is more suitable for large and continuous state and action spaces. Based on actor-critic structures, this method uses DNNs as function approximators to specify deterministic policies that can map large discrete or continuous states into continuous actions \cite{akbari2021age}. In the DPG method, there are the actor (policy $\pi$) and critic (value function $Q$) networks with parameters $\theta^\mu$ and $\theta^Q$, respectively, as well as two copies of actor and critic are denoted by parameters $\theta^{\mu'}$ and $\theta^{Q'}$, respectively \cite{hafner2011reinforcement}. The $Q$ function is updated using temporal-difference methods, similar to DQN. The policy gradient algorithm is applied to update the actor's value through the value from the critic. In this approach, we consider $s_t=o_t$. The state's return is calculated as the sum of discounted future reward as follows:
\begin{align}
	R_t=\sum_{i=t}^T\gamma^{\left(i-t\right)}r\left(s_i,a_i\right).
\end{align}
\begin{algorithm}[h!]
	\small
	\renewcommand{\arraystretch}{0.4}
	\caption{DDPG Algorithm \cite{lillicrap2015continuous}}
	\label{DDPG}
	\textbf{Input:} Initialize actor $\mu\left(s|\theta^\mu\right)$ and critic network $Q\left(s,a|\theta^Q\right)$ with weights $\theta^\mu$ and $\theta^Q$, with random value. 
	\\\textbf{Input:} Initialize target network $\mu'$ and $Q'$, with weights $\theta^{\mu'}\leftarrow\theta^{\mu}$, $\theta^{Q'}\leftarrow\theta^{Q}$
	\\\textbf{Input:} Initialize the replay buffer $\mathcal{B}$
	\\\For{episodes=1 to E}{
		Initialize an action exploration process using a random process $\epsilon$
		\\Receive the initial observation state $s_1$
		\\\For{t=1 to $T$}{
			Based on the current policy and exploration noise at the time slot $t$, take action $a_t=\mu\left(s_t|\theta^\mu\right)+\epsilon_t$
			\\Take action $a_t$ and receive reward $r_t$ and observe new state $s_{t+1}$
			\\Store transition $\left(s_t, a_t, r_t, s_{t+1}\right)$ in $\mathcal{B}$
			\\Sample a random minibatch of $N$ transitions $\left(s_i, a_i, r_i, s_{i+1}\right)$ from $\mathcal{B}$
			\\Set $y_i=r_i+\gamma Q'\left(s_{i+1},\mu'\left(s_{i+1}\big{|}\theta^{\mu'}\right)\Big{|}\theta^{Q'}\right)$
			\\Minimize the loss function to update the critic:
			\begin{align}
				&\nonumber L=\frac{1}{N}\sum_{i}\left(y_i-Q\left(s_i,a_i\big{|}\theta^Q\right)\right)^2
			\end{align}
			\\Using the sampled policy gradient, update the actor policy:
			\begin{align}
				&\nonumber \nabla_{\theta^\mu}J\approx
				\\ \nonumber
				&\frac{1}{N}\sum_{i}\nabla_{a}Q\left(s,a\big{|}\theta^Q\right)\Big{|}_{s=s_i, a=\mu\left(s_i\right)}\nabla_{\theta^\mu}\mu\left(s\big{|}\theta^\mu\right)\big{|}_{s_i}
			\end{align}
			Update the target network parameters:
			\begin{align}
				&\nonumber \theta^{Q'}\leftarrow\tau\theta^{Q}+(1-\tau)\theta^{Q'}\\
				&\nonumber \theta^{\mu'}\leftarrow\tau\theta^{\mu}+(1-\tau)\theta^{\mu'}
			\end{align}
		} 
	}
\end{algorithm}
It is essential to keep in mind that the return depends on the actions taken, and thus on the policy $\pi$, and may be stochastic. In RL, the purpose is to learn a policy that maximizes the expected return from the start distribution $J=\mathbb{E}_{r_i,s_i\sim E,a_i\sim\pi}[R_1]$. The Bellman equation is employed to learn the action-value function $Q(s,a|\theta^Q)$ as in the DQN approach; therefore, we have:
\begin{align}\label{Q}
	&Q^{\pi}\left(s_t,a_t\right)=
	\\ \nonumber
	&\mathbb{E}_{r_t,s_{t+1}\sim E}\left[r\left(s_t,a_t\right)+\gamma\mathbb{E}_{a_{t+1}\sim\pi}\left[Q^\pi\left(s_{t+1},a_{t+1}\right)\right]\right].
\end{align}
Here, the target policy is deterministic; we can define it as a function $\mu: \mathcal S\leftarrow\mathcal A$ and ignore the inner expectation of \eqref{Q}. The expectation is only affected by the environment. In other words, the action-value function can be learned using transitions generated from a different stochastic behavior policy $\varkappa$. We assume function approximators parameterized by $\theta^Q$, which we optimize by minimizing the loss function as follows:
\begin{align}
	&L\left(\theta^Q\right)=\mathbb{E}_{s_t\sim\rho^{\varkappa},a_t\sim\varkappa,r_t\sim E}\left[\left(Q\left(s_t,a_t|\theta^Q\right)-y_t\right)^2\right].
\end{align}
where $y_t$ is:
\begin{align}
	&y_t=r\left(s_t,a_t\right)+\gamma Q\left(s_{t+1},\mu\left(s_{t+1}\right)|\theta^Q\right),
\end{align}
Although $y_t$ is likewise dependent on $\theta^Q$, this is usually ignored. The parameterized actor function $\mu(s|\theta^\mu)$ is maintained in the DPG method, which is used to specify the current policy. The Bellman equation is applied to learn the critic $Q(s,a)$ as in the Q-learning approach. To update the actor, a chain rule is applied to the expected return from the start distribution $J$ based on actor parameters; therefore, we can write:
\begin{align}
	&\nabla_{\theta^\mu}J\approx\mathbb{E}_{s_t\sim\rho^\varkappa}\left[\nabla_{\theta^\mu}Q\left(s,a|\theta^Q\right)\big{|}_{s=s_t,a=\mu\left(s_t|\theta^\mu\right)}\right]=
	\\ \nonumber
	&\mathbb{E}_{s_t\sim\rho^{\varkappa}}\left[\nabla_{a}Q\left(s,a|\theta^Q\right)\big{|}_{s=s_t,a=\mu\left(s_t\right)}\nabla_{\theta_\mu}\mu\left(s|\theta^\mu\right)\big{|}_{s=s_t}\right].
\end{align}
Since samples are not independently and identically distributed in most optimization algorithms, the DDPG method uses a replay buffer $\mathcal{B}$ to address this challenge similar to DQN. In the replay buffer, the cache size is limited. Transitions are sampled from the environment based on the exploration policy, and the tuple $(s_t,a_t,r_t,s_{t+1})$ is collected in the replay buffer. The oldest samples are removed from the replay buffer when it is full. The actor and critic update using sampling a minibatch uniformly of the buffer in every time slot. To compute the target values, we create copies of the actor and critic networks, $Q'(s,a|Q^{\theta'})$ and $\mu'(s|\theta^{\mu'})$, respectively. By tracking the learned networks slowly, the target networks' weights can be updated: $\theta'\leftarrow\tau\theta+(1-\tau)\theta'$ with $\tau\ll1$. As a result, the learning stability is improving because the target values can only change slowly. A noise sampled from a noise process $\epsilon$ is added to actor policy to build an exploration policy $\mu'$; therefore, we have:
\begin{align}
	&\mu'\left(s_t\right)=\mu\left(s_t|\theta^{\mu}_t\right)+\epsilon,
\end{align}
where $\epsilon$ can be selected according to the environment. The complete pseudo-code of the DDPG approach shows in Algorithm \ref{DDPG}.
\subsection{SAC Algorithm}
The SAC is the off-policy actor-critic DRL algorithm based on the maximum entropy RL framework for continuous action spaces. In other words, the optimal policy of this approach is to maximize its entropy-regularized reward rather than to maximize the discounted cumulative reward. According to this framework, the actor wants to maximize the expected reward in addition to maximize entropy. In this method, the infinite-horizon MDP is defined by the tuple $(\mathcal S,\mathcal A,p,r)$ in which the state space $\mathcal S$ and the action space $\mathcal A$ are continuous, and the unknown state transition probability $p:\mathcal S\times\mathcal S\times\mathcal A\rightarrow[0,\infty)$ indicates the probability density of the next state $s_{t+1}$ based on the current state $s_t$ and action $a_t$. We employ $\rho_{\pi}(s_t)$ and $\rho_{\pi}(s_t,a_t)$ to describe the state and state-action marginals of the trajectory distribution induced by a policy $\pi(a_t|s_t)$. To maximize the entropy, the following formula is considered:
\begin{align}
	J\left(\pi\right)=\sum_{t=0}^{T}\mathbb{E}_{\left(s_t,a_t\right)\sim\rho_{\pi}}\left[r\left(s_t,a_t\right)+\varphi\mathcal H\left(\pi\left(\cdot|s_t\right)\right)\right],
\end{align}
where $\varphi$ is a regularization coefficient. To ensure that entropies and the sum of expected rewards are finite, the objective using discount factor $\gamma$ is extended to infinite horizon problems. Soft policy iteration is extended by the SAC method to the setting with function approximation. In order to improve the policy, the SAC employs a different optimization on both the policy and the value function, rather than estimating the true Q value of the policy $\pi$. We assume a parametrized Q function $Q_{\phi}(s,a)$ and policy $\pi_{\theta}$. Moreover, a target Q network is defined as $Q_{\tilde{\phi}}$, where parameter $\tilde{\phi}$ is computed as an exponentially moving average of $\phi$. Through minimizing the soft Bellman residual, the Q function can be learned as follows:
\begin{align}
	&J_Q\left(\phi\right)=
	\\ \nonumber &\mathbb{E}\left[\left(Q\left(s_t,a_t\right)-r\left(s_t,a_t\right)-\gamma\mathbb{E}_{s_{t+1}}\left[V_{\tilde{\phi}}\left(s_{t+1}\right)\right]\right)^2\right],
\end{align}

\begin{algorithm}[h!]
	\small
	\renewcommand{\arraystretch}{0.4}
	\caption{SAC Algorithm \cite{haarnoja2018soft}}
	\label{SAC}
	\textbf{Hyperparameters:} Step sizes $\lambda_{\pi}, \lambda_{Q}, \lambda_{\varphi}$, target entropy $\mathfrak{e}$, exponentially moving average coefficient $\tau$ 
	\\\textbf{Input:} Initial Q value function parameters $\phi_1$ and $\phi_2$
	\\\textbf{Input:} Initial policy parameters $\theta$
	\\$\mathcal{B}=\emptyset; \tilde{\phi}=\phi,$ for $i\in\{1,2\}$
	\\\For{each iteration}{
		\For{each environment step}{
			$\nonumber a_t\sim\pi_{\theta}\left(\cdot|s_t\right)$\\
			$\nonumber s_{t+1}\sim p\left(s_{t+1}|s_t,a_t\right)$\\
			$\nonumber \mathcal{B}\leftarrow\mathcal{B}\cup\left\{s_t,a_t,r\left(s_t,a_t\right),s_{t+1}\right\}$
		}
		\For{each gradient step}{
			$\nonumber \theta\leftarrow\theta-\lambda_{\pi}\nabla_{\theta}J_{\pi}\left(\theta\right)$\\
			$\nonumber \phi_{i}\leftarrow\phi_i-\lambda_{Q}\nabla J_{Q}\left(\phi_i\right)$ for $i\in\{1,2\}$\\
			$\nonumber \varphi\leftarrow\varphi-\lambda_{\varphi}\nabla J\left(\varphi\right)$\\
			$\nonumber \tilde{\phi}_i\leftarrow\tau\tilde{\phi}_i+\left(1-\tau\right)\phi_i$ for $i\in\{1,2\}$	    	      
		}
	}	                      
\end{algorithm}
where $V_{\tilde{\phi}}(s)$ is:
\begin{align}
	&V_{\tilde{\phi}}\left(s\right)=\mathbb{E}_{\pi_{\theta}}\left[Q_{\tilde{\phi}}\left(s,a\right)-\varphi\log\pi_{\theta}\left(a|s\right)\right].
\end{align}
Furthermore, by minimizing the expected KL-divergence, we can learn policy $\pi_{\theta}$:
\begin{align}\label{J}
	&J_{\pi}\left(\theta\right)=\mathbb{E}_{s\sim\mathcal{B}}\left[\mathbb{E}_{a\sim\pi_{\theta}}\left[\varphi\log\pi_{\theta}\left(a|s\right)-Q_{\phi}\left(s,a\right)\right]\right],
\end{align}
where $\mathcal{B}$ represents the set of previous sampled states and actions or the replay buffer. To reduce the biased Q value problem, the SAC employs two Q-networks (also two target Q-networks), i.e. $Q_{\phi}(s,a)=\min(Q_{\phi_1}(s,a),Q_{\phi_2}(s,a))$. It is possible to optimize $J_{\pi}(\theta)$ in several different ways. A likelihood ratio gradient estimator \cite{williams1992simple} is a common solution for policy gradient approaches because it does not need to use backpropagating the gradient through the target density networks and the policy. However, the target density in the SAC is the Q-function described by a neural network, and we can use the reparameterization trick for the policy network, which often leads to a lower variance estimator. To this end, we reparameterize the policy $\pi_{\theta}$ utilizing a neural network transformation, that gets both the state $s$ and noise vector $\epsilon$ as an input, as follows:
\begin{align}\label{a}
	a=f_{\theta}\left(s,\epsilon\right).
\end{align}
Therefore, based on \eqref{a}, we can rewrite \eqref{J} as follows:
\begin{align}
	J_{\pi}\left(\theta\right)=\mathbb{E}_{s\sim\mathcal{B},\epsilon\sim\mathcal N}\left[\varphi\log\pi_{\theta}\left(f_{\theta}\left(s,\epsilon\right)|s\right)-Q_{\phi}\left(s,f_{\theta}\left(s,\epsilon\right)\right)\right],
\end{align}
where $\mathcal N$ is a standard Gaussian distribution, and $\pi_{\theta}$ is defined in terms of $f_{\theta}$ implicitly. Lastly, the SAC gives a method to automatically update the regularization coefficient $\varphi$ through minimizing the loss function as follows:
\begin{align}
	J\left(\varphi\right)=\mathbb{E}_{a\sim\pi_{\theta}}\left[-\varphi\log\pi_{\theta}\left(a|s\right)-\varphi\mathfrak{e}\right],
\end{align}
where $\mathfrak{e}$ is a hyperparameter that represents the target entropy. In Algorithm \ref{SAC}, a complete description of the SAC method is given.         
\subsection{Distributed Algorithm}
In this method, we consider two agents for the SAC algorithm, one for the RAN domain and the other for the core domain. These two agents have no interaction with each other. At first, the radio agent solves the RAN problem, and then the core agent solves the core problem. In other words, the main difference between the distributed strategy and the multi-agent approach is that in the multi-agent method, agents interact with each other, but not in the distributed way.
\section{Computational Complexity and Convergence Analysis}\label{VI}
Computational complexity and convergence are two essential criteria in solving optimization problems. On the one hand, in DRL algorithms, one method may have a high computational complexity, which increases the run time of the simulation. But it is possible; it has better performance than other methods and has a faster convergence rate. To better understand the subject, more details are provided in the following two subsections.
\subsection{Computational Complexity}
In this subsection, we investigate the computational complexity of the main method and the baselines. The computational complexity of the proposed system model is one of the significant and practical factors. Therefore, we examine two aspects of the computational complexity of our proposed system model: the action selection process and the training process. 
\\To compute the computational complexity, we need to define several parameters. For this purpose, we define the total number of episodes by $N_{\text{Episod}}$, the total number of neural network layers by $Z$, the total number of hidden layers by $H$, and the total number of neurons in each layer by $Y$. Accordingly, the number of neurons in the $z$-th layer is denoted by $Y_z$. Also, $|S|$ and $|A|$ are the size of the total state and action spaces, respectively.
\\Moreover, for the distributed approach, we define state and action of RAN domain by $S_{\text{RAN}}$ and $A_{\text{RAN}}$, respectively, and for the core domain, we define state and action by $S_{\text{Core}}$ and $A_{\text{Core}}$, respectively.
\subsubsection{Computational Complexity of Action Selection}
Accordingly, the computational complexity of neural network-based algorithms depends on the network structure and its layers. The complexity of back-propagation on a fully connected neural network depends on the multiplication of the input, hidden layers and, output. The RDPG method consists of one actor and one critic neural network. Due to the use of $L$ previous trajectories for action selection, the production of the sizes of each two consecutive layers of actor and critic can be computed by \eqref{actor}, \eqref{critic}, respectively.
\begin{align}\label{actor}
	\underbrace{L\times\left(\left|S\right|+\left|A\right|\right)\times H}_{\text{Layer}\,1},\dots,\underbrace{H^2}_{\text{Layer}\,h},\dots,\underbrace{H\times\left|A\right|}_{\text{Layer}\,H},
\end{align}
\begin{align}\label{critic}
	\underbrace{L\times\left(\left|S\right|+2\times\left|A\right|\right)\times H}_{\text{Layer}\,1},\dots,\underbrace{H^2}_{\text{Layer}\,h},\dots,\underbrace{H\times\left|A\right|}_{\text{Layer}\,H}.
\end{align}
Hence, in the RDPG algorithm, the computational complexity of action selection is $\mathcal{O}(H^2)$. 
\subsubsection{Computational Complexity of Training}
The training process complexity in the RDPG approach with $H$ hidden layers with $Y$ neurons is calculated as follows \cite{sheikhzadeh2021ai}:
\begin{align}
	\mathcal{O}\left(\mathfrak{B}L\times\left(|S|+2\times|A|\right)Y^H\right),
\end{align}
where $\mathfrak{B}$ is the size of the training batch. In addition, to better understand the differences between the solutions used, Table \ref{complexity} compares the computational complexity of the RDPG method with the baselines.
\begin{table}[h!]
	\centering
	\caption{Comparison of the computational complexity of RDPG method with baselines}
	\label{complexity}
	\scalebox{1}{
	\begin{tabular}{|c|c|}
		\hline
				\rowcolor[HTML]{38FFF8} 
		\textbf{Algorithm} & \textbf{Computational complexity}
		 \\ \hline
		RDPG               & $\mathcal{O}\left(\mathfrak{B}L\times\left(|S|+2\times|A|\right)Y^H\right)$                  \\ \hline
		DDPG               & $\mathcal{O}\left(\left(\sum_{z=0}^{Z-1}Y_{z}\times Y_{z+1}\right)\times\mathfrak{B}\times N_{\text{Episod}}\right)$                   \\ \hline
		SAC                & $\mathcal{O}\left(\left(\sum_{z=0}^{Z-1}Y_{z}\times Y_{z+1}\right)\times\mathfrak{B}\times N_{\text{Episod}}\right)$                   \\ \hline
		Distributed        & \begin{tabular}[c]{@{}c@{}}$\mathcal{O}\Big{(}\mathcal{O}\left(|S_{\text{RAN}}|\times Y_{2}+Y_{2}\times Y_{3}+Y_3\times|A_{\text{RAN}}|\right),$\\ 
		$\mathcal{O}\left(|S_{\text{Core}}|\times Y_{2}+Y_{2}\times Y_{3}+Y_3\times|A_{\text{Core}}|\right)\Big{)}$\end{tabular}                   \\ \hline
		Greedy             & $\mathcal{O}\left(I+|\mathcal{S}|\times\left(|\mathcal{N}|\log_{2}^{|\mathcal{N}|}+|\mathcal{L}|\right)+\mathcal{S}\times F\right)$                   \\ \hline
	\end{tabular}
}
\end{table}
\subsection{Convergence Analysis}
\begin{figure}[h!]
	\centering
	\includegraphics[width=\linewidth]{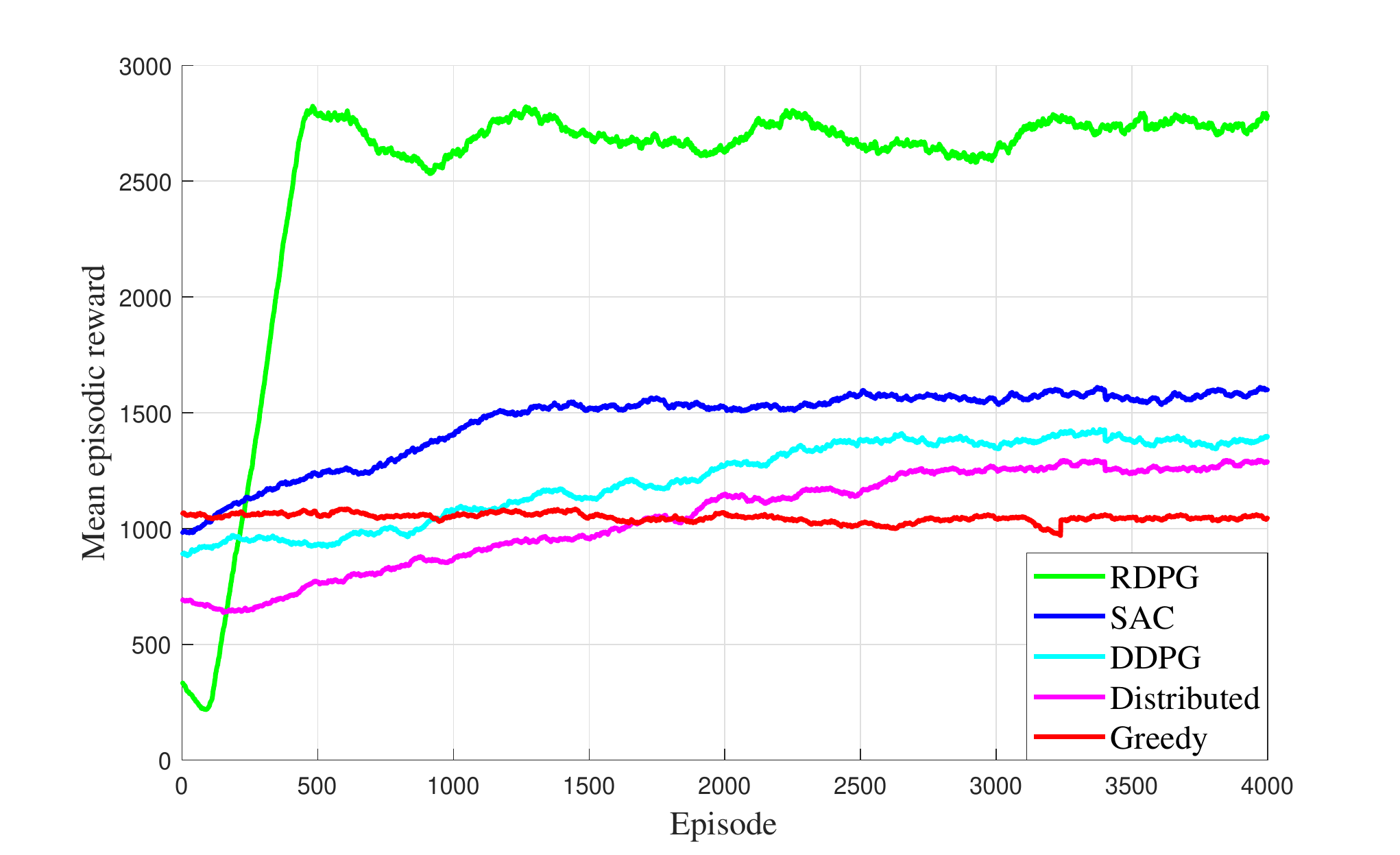}
	\caption{Mean episodic reward versus episode}
	\label{fig:convergence}
\end{figure}

In this subsection, we examine the convergence of the RDPG algorithm and other approaches. In the Q-learning algorithm, the Q-function can converge to the optimal Q-function as $t\rightarrow\infty$ with probability 1, if actor and critic network learning rate $\alpha'$ and $\alpha''$ are deterministic, non-increasing, and satisfy the following formulas \cite{grondman2012survey}:
\begin{align}
	&\sum_{t=0}^{\infty}\alpha''_{t}=\infty,\quad\sum_{t=0}^{\infty}\left(\alpha''_{t}\right)^2<\infty,
\end{align}
\begin{align}
	&\sum_{t=0}^{\infty}\alpha'_{t}=\infty,\quad\sum_{t=0}^{\infty}\left(\alpha'_{t}\right)^2<\infty,\quad\lim\limits_{t\rightarrow\infty}\frac{\alpha'}{\alpha''}=0.
\end{align}
Also $|r_{t}(s_t,a_t)|$ be bounded \cite{watkins1992q}. We employ an inverse time decaying learning rate to achieve fast convergence and effectively train DNN; that in the early episodes, it utilizes the large learning rate to avoid the network from getting trapped in a bad local optimum. Moreover, to converge to a good local optimum, it applies the small learning rate in the last training episodes \cite{you2019does}. Fig. \ref{fig:convergence} shows the mean episodic reward versus episode. This figure shows the simulation results for 20 users when we set the value of CSI and demand uncertainty bound to be 5$\%$ and 10$\%$, respectively. As can be seen, the RDPG method converges faster than other methods in terms of convergence. Given that the RDPG is memory-based, it does not have good results in the early episodes because the history contains little information and is not enough. But, over time and increasing episodes, the agent gets better rewards by exploiting the history. Moreover, in the DRL algorithms that we use (i.e., RDPG, SAC, DDPG, and distributed), action with noise is selected. Hence in the early episodes, the greedy way is better than other methods. But as the number of episodes increases, the performance of this method worsens compared to other approaches.

\section{Simulation Results}\label{VII}
In this section, we first introduce the simulation environment, then we examine the simulation results in terms of different aspects between the RDPG approach and other methods.
\subsection{Simulation Environment}
Here, we describe the simulation environment used for evaluating the efficiency of employed algorithms. To simulate the proposed problem, we use the programming language, \textit{Python 3.8.12}, and compiler, \textit{Spyder 5.1.5}. Simulations are run on a personal computer with 8 cores, 3.8GHz Intel Core i7-10700K CPU, and 16GB RAM. In the RAN domain, the users are uniformly distributed in a square area 1000m×1000m with 4 BSs where the maximum power of each BS is 4 watts (36.02 dBm). We consider 10 subchannels with the frequency bandwidth of 200 kHz, and the minimum data rate of the downlink is equal to 1 bps/Hz. In the core domain, we use the \textit{Abilene} network topology with 12 nodes and 15 links from \textit{SNDlib}\footnote{\textit{SNDlib} \cite{sndlib} is a library of test cases for the design of survivable fixed telecommunication networks.} \cite{SNDlib10,OrlowskiPioroTomaszewskiWessaely2010}, that we used the \textit{NetworkX} libraries\footnote{\textit{NetworkX} \cite{networkx} is a \textit{Python} package for building, manipulating, and studying complex networks' structure, dynamics, and functions.} \cite{hagberg2013networkx} in \textit{Python} to implement the graph $\mathcal G=(\mathcal N,\mathcal L)$ similar to the existing work \cite{fu2019dynamic}. \textit{Tensorflow} and \textit{PyTorch} libraries\footnote{\textit{TensorFlow} and \textit{PyTorch} are open-source ML libraries used to develop and train neural network-based deep learning models.} in \textit{Python} with \textit{Adam}\footnote{Adam is an alternative optimization method to stochastic gradient descent for training deep learning algorithms.} optimizer are applied to implement the DNN. We used \textit{Tensorflow 2.6.1} and \textit{Torch 1.4.0} for simulation. At the start of the simulation, we randomly select several nodes from the network nodes to set the ingress and egress nodes for the set of slices $\mathcal S$. The proposed system model has 6 VNFs where each VNF needs a different processing time. Furthermore, each node can host at most 6 VMs, and each VM can host a maximum of 6 VNFs. We set capacities of bandwidth and memory of each physical link and node to 1 Gbps and 1 Gbyte, respectively. Additionally, CPU, RAM, and storage capacities are set to 1200 CPU cycle/Hz, 1000 Mb, and 1000 Mb, respectively. Moreover, more details of the simulation parameters are listed in Table \ref{Simulation parameters}. To better understand the proposed problem, the source code of the simulation of the main approach and baselines are available in \cite{4jps-kt78-22}.
\begin{table}[h!]
	\centering
	\caption {Simulation parameters}
	\label{Simulation parameters}
	\scalebox{0.7}{
		\begin{tabular}{|c|l|c|}
			\hline
					\rowcolor[HTML]{38FFF8} 
			\textbf{Parameter}      & \multicolumn{1}{c|}{\textbf{Description}}  & \textbf{Value}     \\ \hline
			\multicolumn{3}{|c|}{\textbf{E2E NwS environment}}                                               \\ \hline
			$C$                     & Number of total users                      & 24                 \\ \hline
			$I$                     & Number of total BSs (cells)                & 4 \cite{tong2020communication}                  \\ \hline
			$K$                     & Number of total subchannels                & 10                 \\ \hline
			$S$                     & Number of total slice types                & 3 \cite{dogra2020survey,habibi2019comprehensive}                  \\ \hline
			$B$                     & Total available bandwidth                  & 200 KHz            \\ \hline
			$B_k$                   & Bandwidth of each subchannel               & 20 KHz             \\ \hline
			$h^{i,k}_{m_s}$         & Channel gain                               & Rayleigh fading \cite{korrai2020joint}    \\ \hline
			$\sigma^{2}$    & Power of AWGN                                      & -174 dBm/Hz \cite{moltafet2019robust}        \\ \hline
			$P^i_{\text{max}}$      & Maximum transmitted power by each BS       & 4 Watt (36.02 dBm) \\ \hline
			$\Gamma^{i,k}_{m_s}$    & CSI uncertainty bound (to percentage)      & 0, 2, 4, 6, 8, 10 \cite{korrai2020joint,moltafet2019robust} \\ \hline
			$R^s_{\text{min}}$      & Minimum required data rate of slice $s$    & 1, 1.2, 1.4, 1.6, 1.8, 2, 3, 4, 5 bps/Hz                 \\ \hline
			$\nu$                   & Speed of light in meter per second         & $3\times10^8$ m/s                           \\ \hline
			$N$                     & Number of total nodes                      & 12 \cite{sndlib,SNDlib10,OrlowskiPioroTomaszewskiWessaely2010}               \\ \hline
			$L$                     & Number of total links                      & 15 \cite{sndlib,SNDlib10,OrlowskiPioroTomaszewskiWessaely2010}                                                    \\ \hline
			$V_{\text{Total}}$      & Number of total VMs                        & 6                                       \\ \hline
			$\mathcal F$            & Set of all VNFs types                      & 6 \cite{savi2019impact}                 \\ \hline
			$\tau^{s}_{\text{max}}$ & Maximum tolerable delay time               & 60, 100, 200, 300, 400, 500 ms \cite{savi2019impact}                 \\ \hline
			$\hat{w}_s$             & Demand uncertainty bound (to percentage)   & 0, 5, 10, 15, 20, 25, 30 \cite{gholamipour2021online}     \\ \hline
			$\theta_1$              & Revenue scaling factor                     & 60                                      \\ \hline
			$\theta_2$              & Cost scaling factor                        & 1                                       \\ \hline
			\multicolumn{3}{|c|}{\textbf{Deep neural network}}                                                             \\ \hline
			$N_{\text{Episod}}$     & Number of episodes                         & 4000                                        \\ \hline
			$\mathfrak{B}$          & Batch size                                 & 64                                          \\ \hline
			$\tau$                  & Target network update period               & 0.001                                       \\ \hline
			$H$                     & Number of hidden layers                    & 2                                           \\ \hline
			$Y$                     & Number of neurons in each hidden layer     & 512                                         \\ \hline
			-                       & Activation function in hidden layers       & ReLU \cite{sheikhzadeh2021ai}               \\ \hline
			-                       & Activation function in output layer        & tanh \cite{sheikhzadeh2021ai}               \\ \hline
			$\alpha'$               & Actor network learning rate                & 0.00001                                     \\ \hline
			$\alpha''$               & Critic network learning rate              & 0.00005                                     \\ \hline
			$\gamma$                & Discount factor                            & 0.80                                        \\ \hline
			$\mathcal{B}$              & Replay buffer size                         & 600000                                        \\ \hline
		\end{tabular}
	}
\end{table}
\subsection{Performance of Simulation Results and Metrics }
We analyze the impact of the main parameters, such as the number of users, demand uncertainty, CSI uncertainty, maximum tolerable delay time, and minimum required data rate on different baseline algorithms.
\subsubsection{Effect of Number of Users}
\begin{figure}[h!] 
	\centering
	\includegraphics[width=\linewidth]{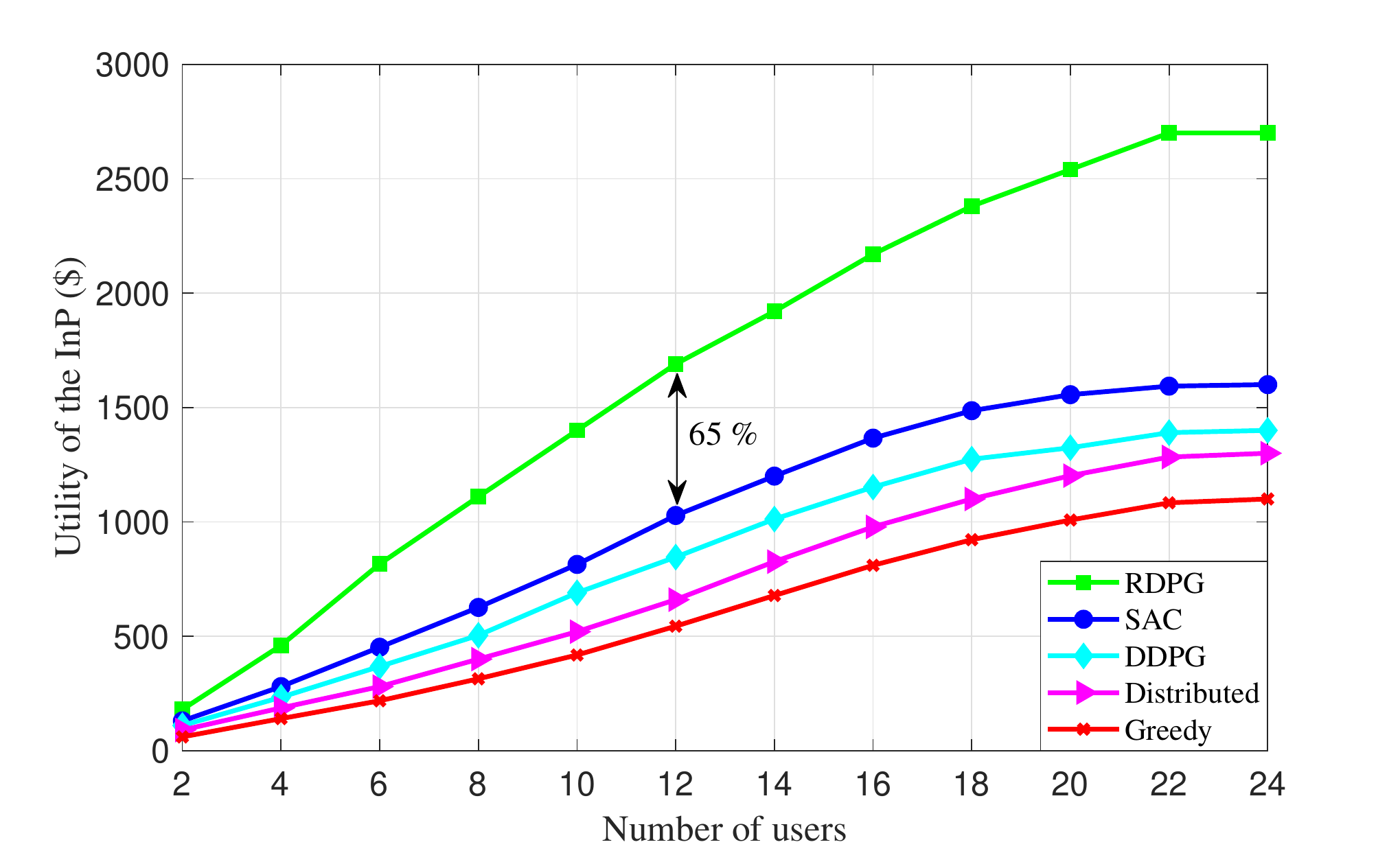}
	\caption{Utility of the InP versus number of users.}
	\label{fig:user}
\end{figure}

In Fig. \ref{fig:user}, the utility of the InP versus number of users is depicted. As the number of users increases, the data rate allocated increases. Therefore, the InP utility increases because InP revenue comes from the data rate. However, due to the limited network resources, after the number of users changes from 22 to 24, InP no longer accepts new users, therefore the utility remains constant. As shown in Figure 4, the RDPG algorithm performs an average of 65$\%$ better than the SAC. In this scenario, we consider the value of CSI and demand uncertainty bound to be 5$\%$ and 10$\%$, respectively.

\subsubsection{Effect of Demand Uncertainty}
\begin{figure}[h!]
	\centering
	\includegraphics[width=\linewidth]{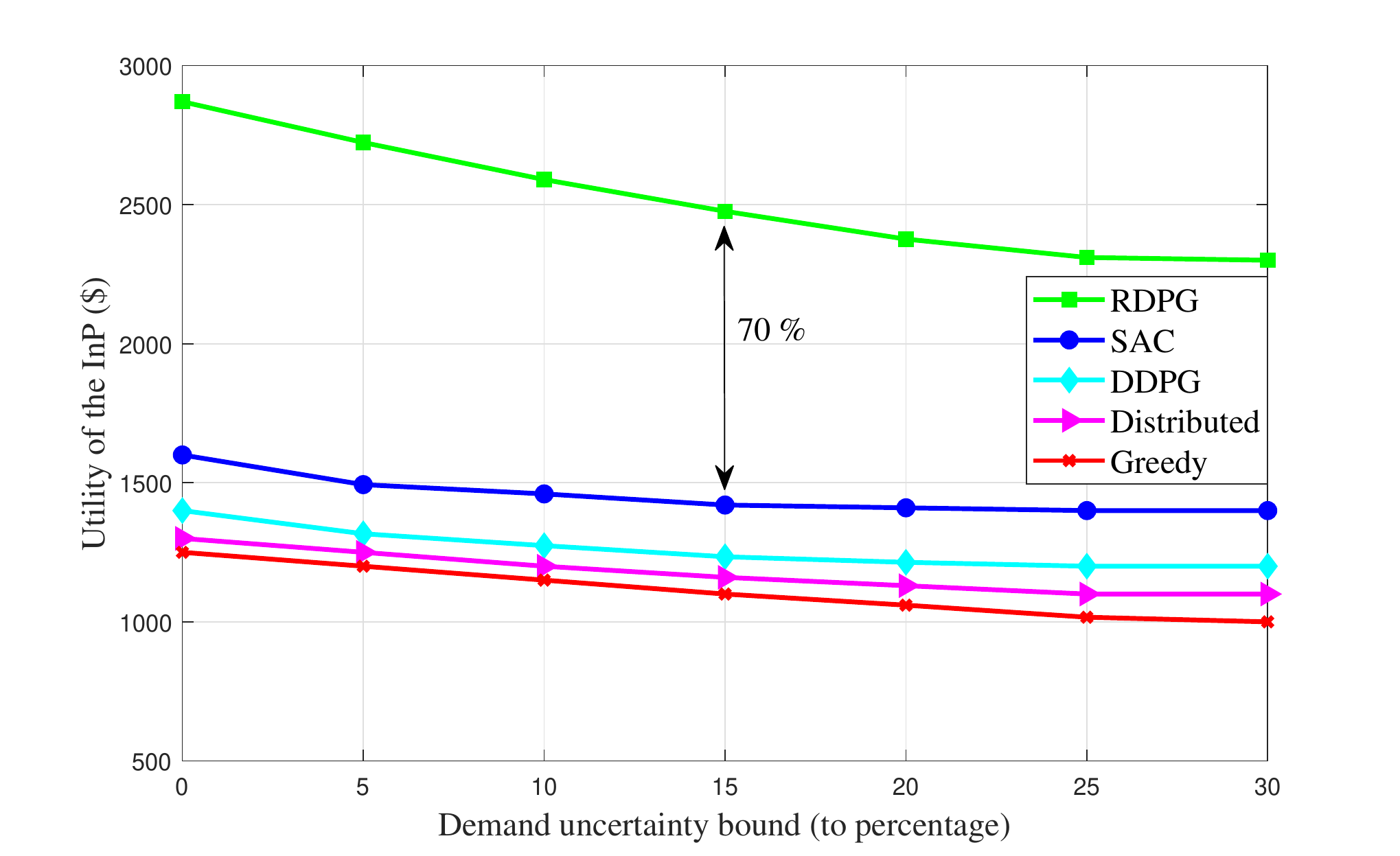}
	\caption{Utility of the InP versus demand uncertainty bound.}
	\label{fig:demand}
\end{figure}

To investigate the effect of demand uncertainty on the objective function of our proposed problem, we keep the value of CSI uncertainty bound $\Gamma^{i,k}_{m_s}$ constant at 2$\%$ and change the value of demand uncertainty bound $\hat{w}_{m_s}$ in the range 0$\%$ to 30$\%$. We only know the expectation and variance of each user's requested data rate, and we do not know the exact amount of data rate requested in each slice. Therefore, we need a history of the average variance of previous user requests. As shown in Fig. \ref{fig:demand}, as the demand uncertainty bound increases, the InP's utility decreases. The RDPG strategy is more powerful than other methods in this scenario. This is because this algorithm has a history and is suitable for problems that include uncertain information, such as our work.

\subsubsection{Effect of CSI Uncertainty}
\begin{figure}[h!]
	\centering
	\includegraphics[width=\linewidth]{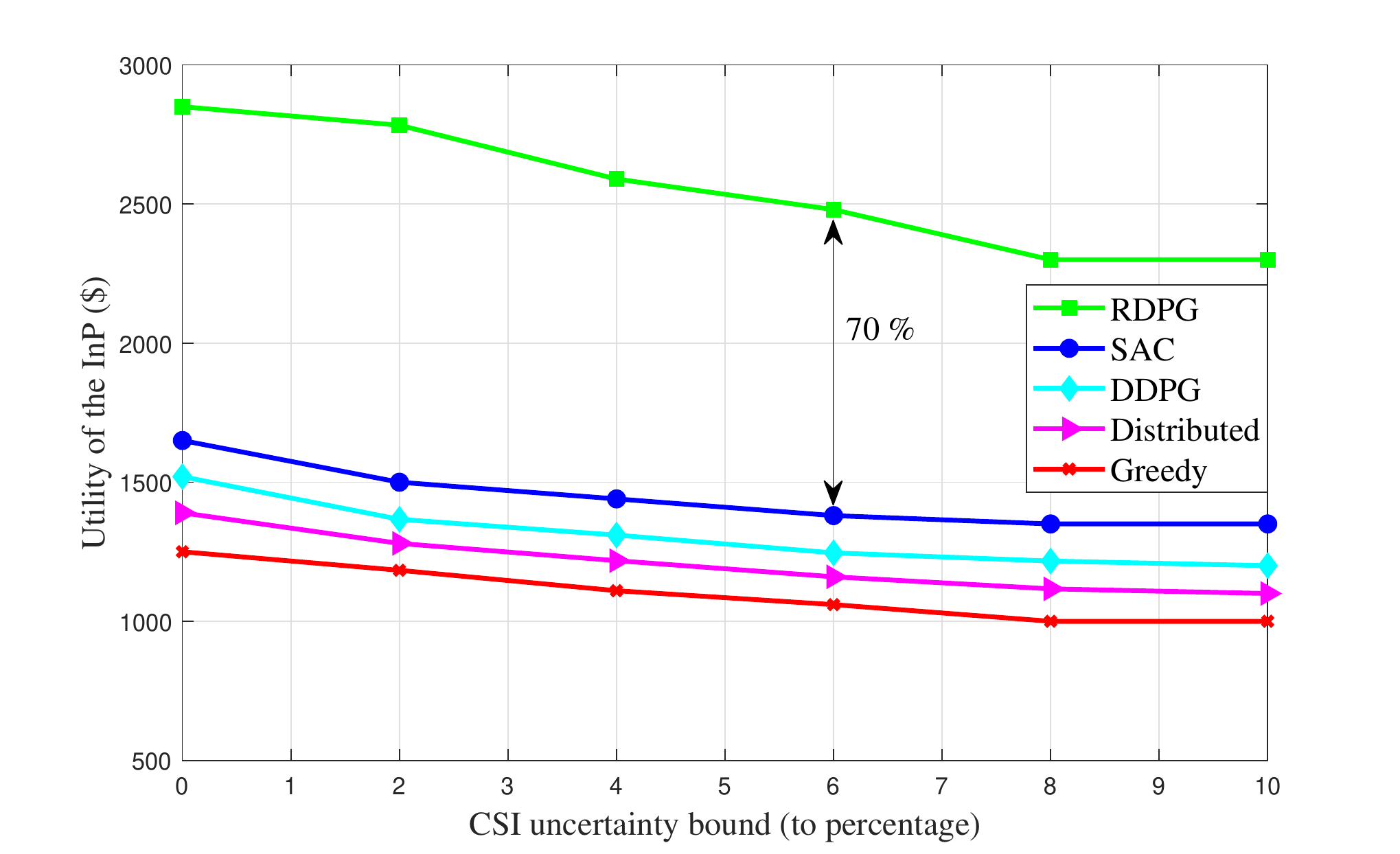}
	\caption{Utility of the InP versus CSI uncertainty bound.}
	\label{fig:csi}
\end{figure}

In this scenario, we keep the value of demand uncertainty bound $\hat{w}_{m_s}$ constant at 5$\%$ and check the InP's utility by increasing the CSI uncertainty bound value $\Gamma^{i,k}_{m_s}$ in a range of 0$\%$ to 10$\%$. The channel gain has a direct impact on the data rate formula. As the CSI uncertainty bound increases, the amount of data rate allocated to the user in the RAN domain decreases. In our proposed problem, the InP revenue is directly related to the data rate, so the revenue decreases, and as a result, the utility decreases. On average, the RDPG method performs 70$\%$ better than the SAC method in this scenario. This is clearly shown in Fig. \ref{fig:csi}. Comparing Fig. \ref{fig:demand} and Fig. \ref{fig:csi}, it can be concluded that CSI uncertainty has a more significant impact on utility than demand uncertainty and has a more destructive effect.

\subsubsection{Effect of Maximum Tolerable Delay Time}
\begin{figure}[h!]
	\centering
	\includegraphics[width=\linewidth]{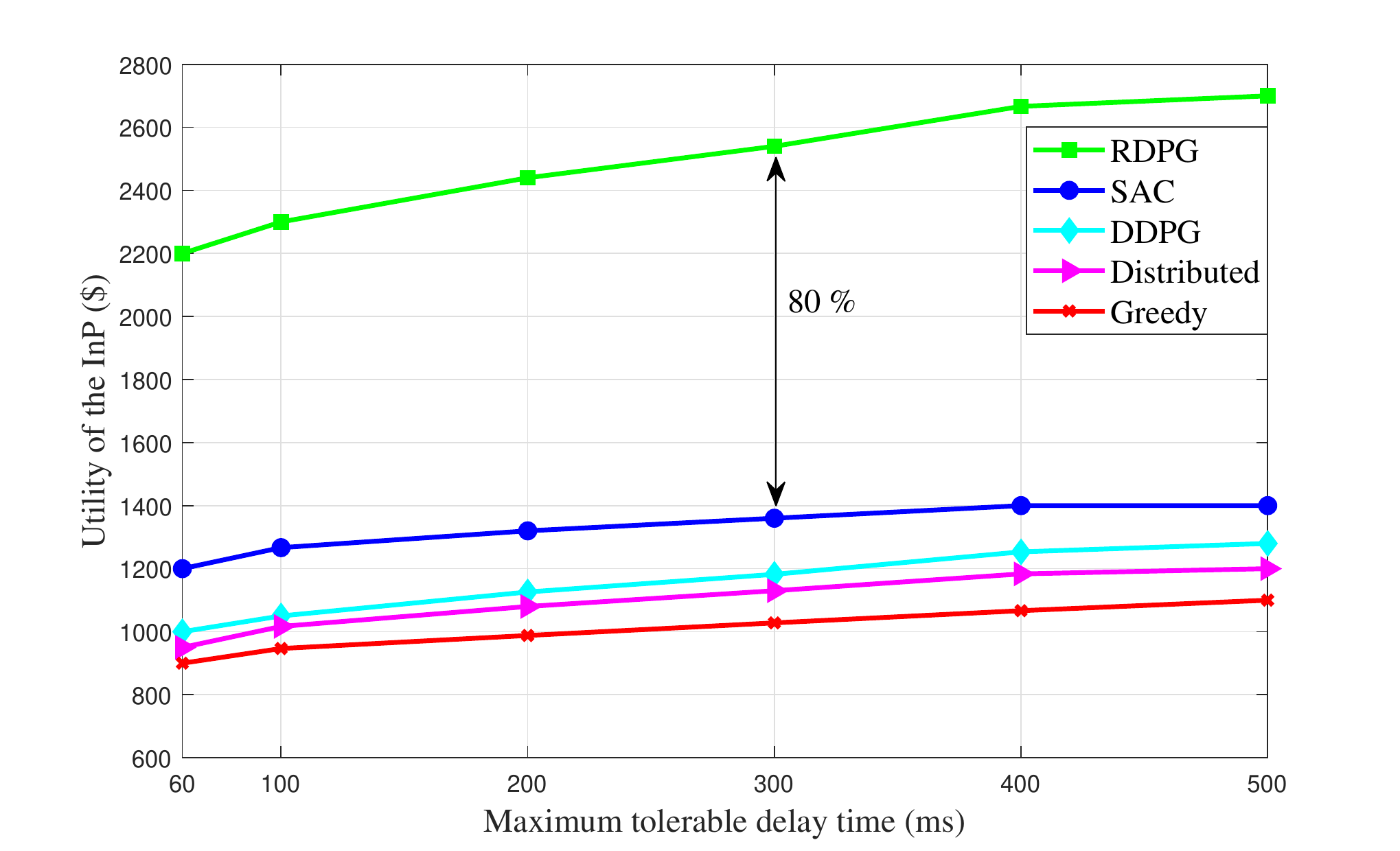}
	\caption{Utility of the InP versus maximum tolerable delay time.}
	\label{fig:mindelay}
\end{figure}

Delay must be guaranteed for each slice. If the InP fails to guarantee a user's expected delay in each slice, that user will not accept, and InP will reject it. In this scenario, CSI and demand uncertainty bound values are fixed to 4$\%$ and 5$\%$, respectively, and we increase the value of maximum tolerable delay time $\tau^{s}_{\text{max}}$ from 60 ms to 500 ms. As shown in Fig. \ref{fig:mindelay}, as the amount of tolerable delay increases, the number of requests accepted by the InP increases; therefore, the utility is increases.

\subsubsection{Effect of Minimum Required Data Rate}
\begin{figure}[h!]
	\centering
	\includegraphics[width=\linewidth]{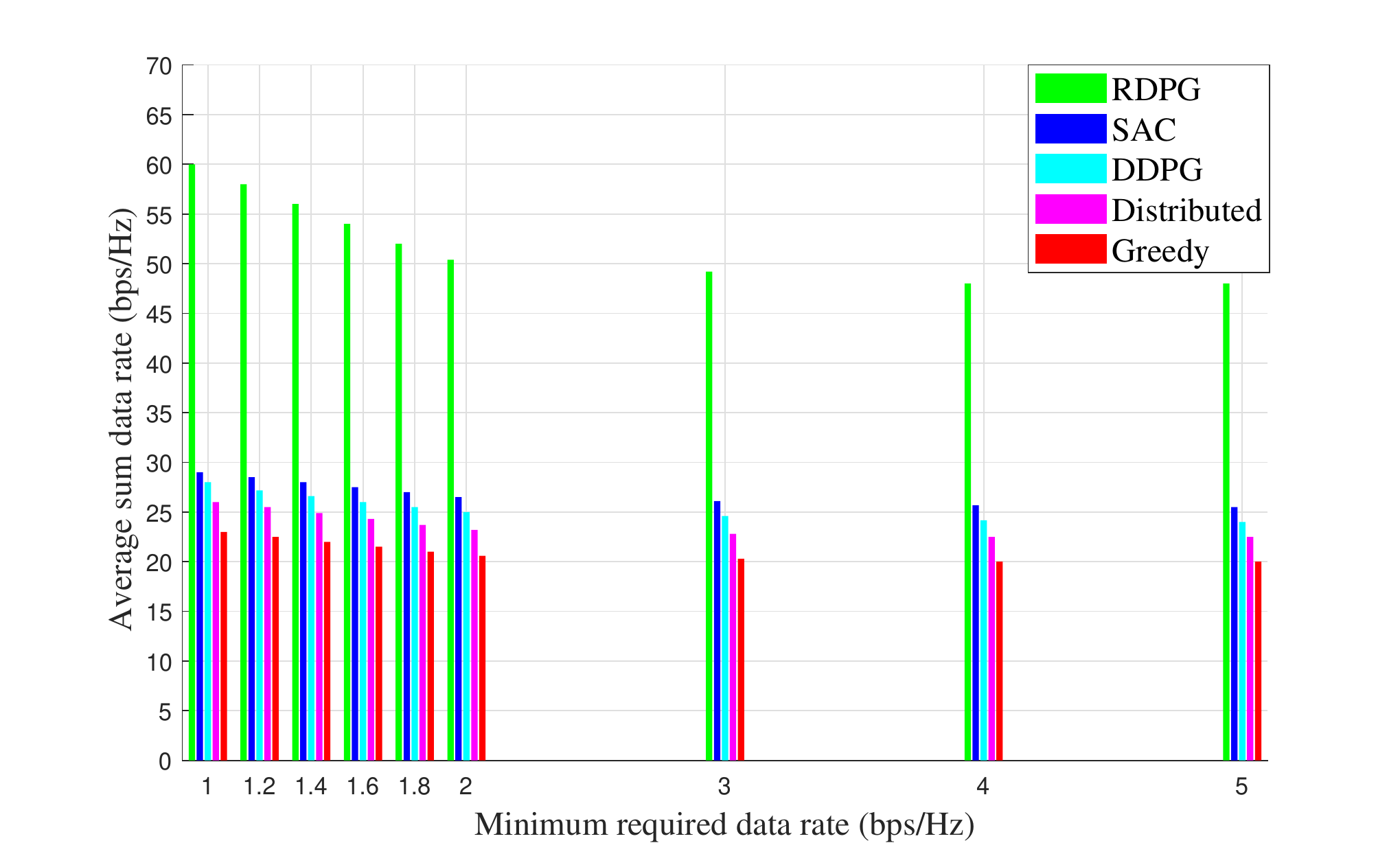}
	\caption{Average sum data rate versus minimum required data rate.}
	\label{fig:rate}
\end{figure}

Here, we limit CSI and demand uncertainty bound to 2$\%$ and 5$\%$, respectively, and then we increase the value of the minimum required data rate $R^s_{\text{min}}$ from 1 bps/Hz to 5 bps/Hz. In Fig. \ref{fig:rate}, the average sum data rate versus minimum data rate required is plotted. As shown, when $ R^s_{\text{min}}$ increases, the average sum data rate decreases. This is because when $R^s_{\text{min}}$ is increased, more subchannels must be assigned to users to satisfy the minimum required data rate, especially for users with poor channel conditions in an extremely deep fade.

\begin{figure}[h!]
	\centering
	\includegraphics[width=\linewidth]{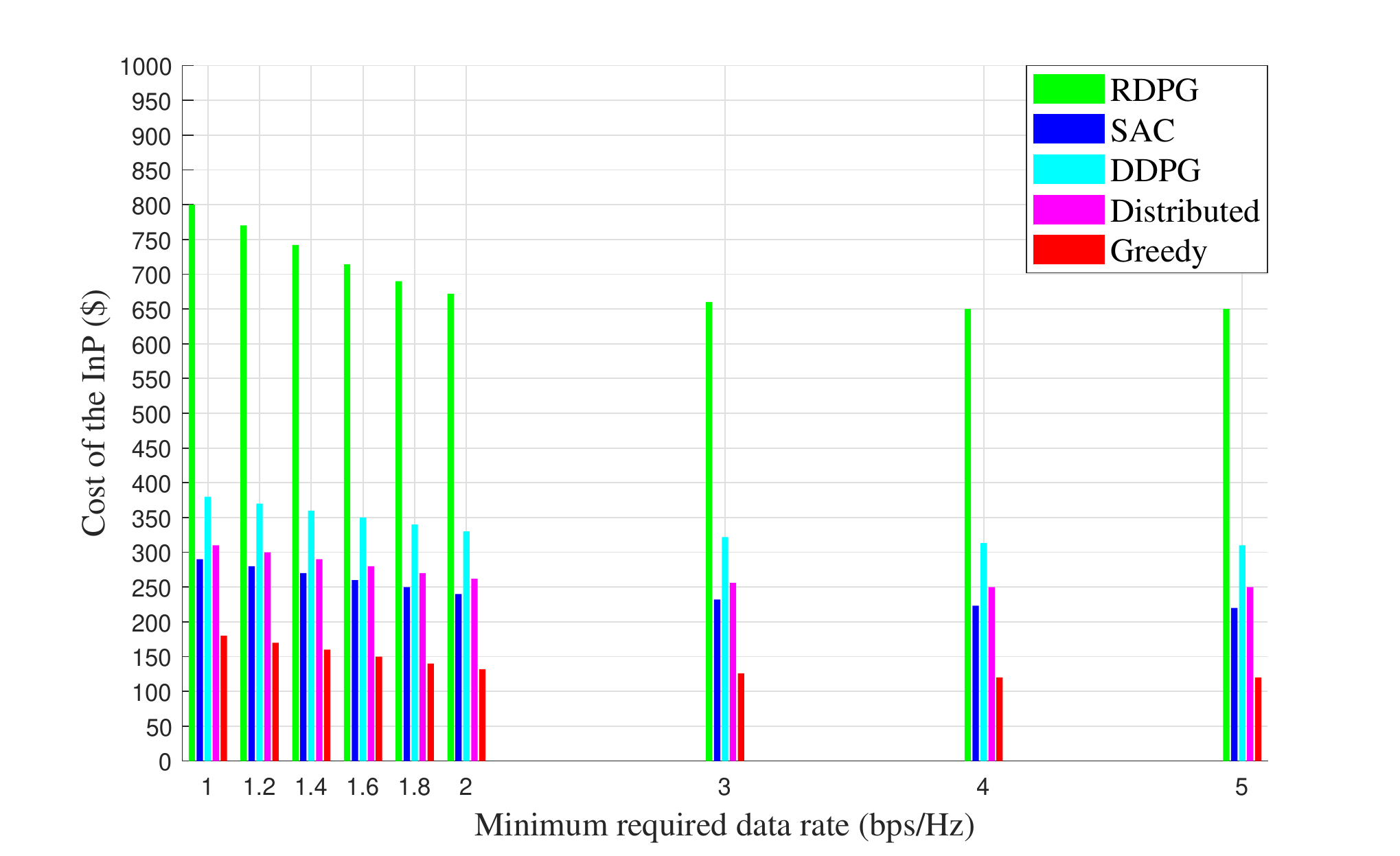}
	\caption{Cost of the InP versus minimum required data rate.}
	\label{fig:cost}
\end{figure}

Additionally, by increasing the value of $R^s_{\text{min}}$, the number of users accepted by InP decreases, therefore the cost of the network decrease. Fig. \ref{fig:cost} shows the details of this issue.

\begin{figure}[h!]
	\centering
	\includegraphics[width=\linewidth]{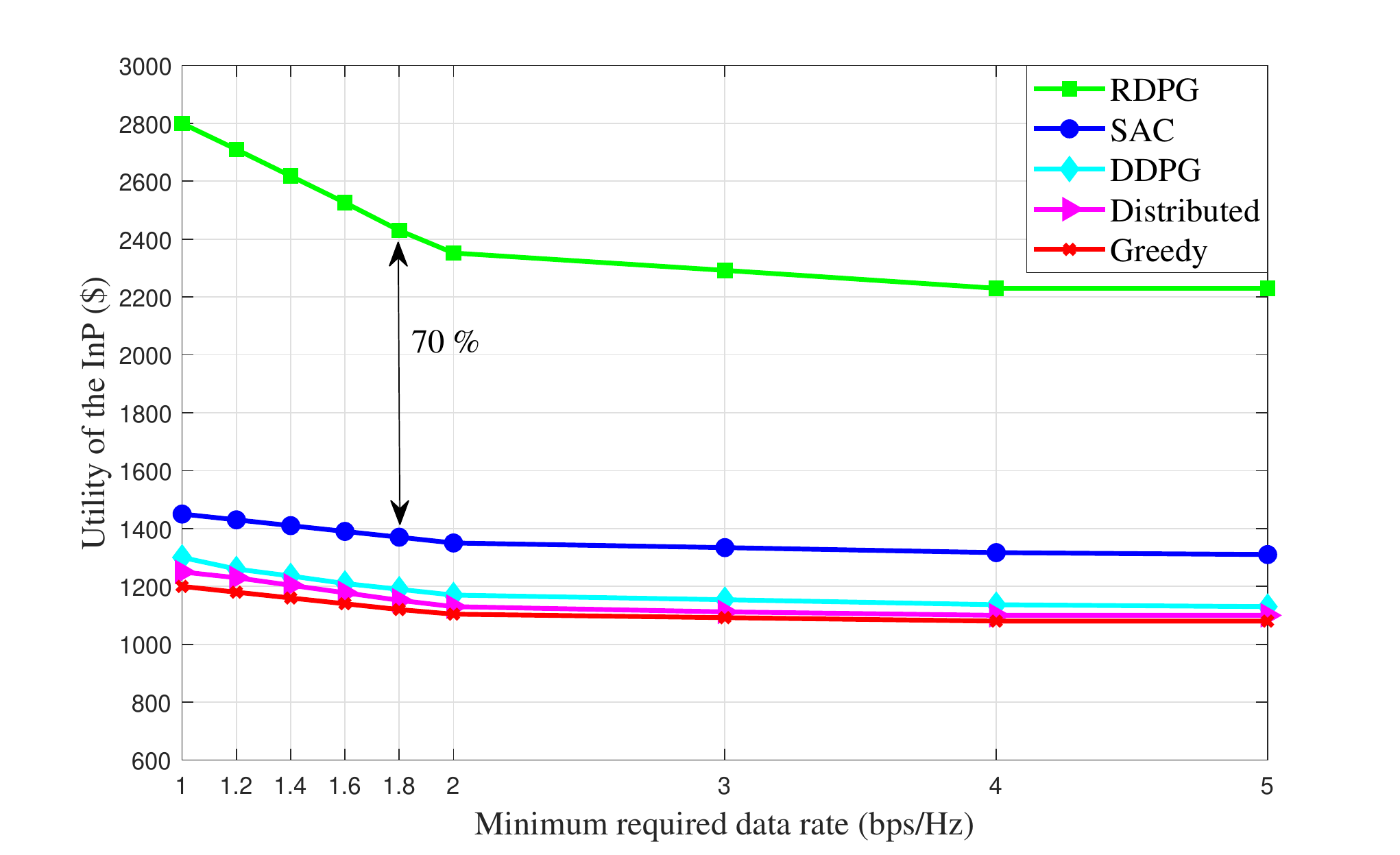}
	\caption{Utility of the InP versus minimum required data rate.}
	\label{fig:utility}
\end{figure}

Finally, as shown in Fig. \ref{fig:utility}, since increasing the $R^s_{\text{min}}$, average sum data, and cost rate is reduced, therefore the utility function, which is the product of the difference between revenue and cost, decreases. To better understand the effect of the minimum required data rate, in addition to using Fig. \ref{fig:rate} and Fig. \ref{fig:cost}, we use Table.\ref{Table: Average sum data rate} and Table.\ref{Table: Cost of the InP} to show the decrease of average sum data rate and cost of the InP, respectively.
\\In this scenario, as in the previous cases, the RDPG algorithm is more robust than the other methods.		
\begin{table*}[h!]
	\centering
	\caption{Average sum data rate versus minimum required data rate.}
	\label{Table: Average sum data rate}
	\scalebox{0.99}{
	\begin{tabular}{|cc|c|c|c|c|c|c|c|c|c|}
		\rowcolor[HTML]{38FFF8}
		\hline
		\multicolumn{2}{|c|}{Minimum required data rate (bps/Hz)}                           & 1  & 1.2  & 1.4  & 1.6  & 1.8  & 2    & 3    & 4    & 5    \\ \hline
		\multicolumn{1}{|c|}{\multirow{5}{*}{Average sum data rate (bps/Hz)}} & RDPG        & 60 & 58   & 56   & 54   & 52   & 50   & 48   & 48   & 48   \\ \cline{2-11} 
		\multicolumn{1}{|c|}{}                                                & SAC         & 29 & 28.5 & 28   & 27.5 & 27   & 26.5 & 26   & 25.5 & 25.5 \\ \cline{2-11} 
		\multicolumn{1}{|c|}{}                                                & DDPG        & 28 & 27   & 26.5 & 26   & 25.5 & 25   & 24.5 & 24   & 24   \\ \cline{2-11} 
		\multicolumn{1}{|c|}{}                                                & Distributed & 26 & 25.5 & 25   & 24.5 & 23.5 & 23   & 22.5 & 22.5 & 22.5 \\ \cline{2-11} 
		\multicolumn{1}{|c|}{}                                                & Greedy      & 23 & 22.5 & 22   & 21.5 & 21   & 20.5 & 20   & 20   & 20   \\ \hline
	\end{tabular}
}
\end{table*}

\begin{table*}[h!]
	\centering
	\caption{Cost of the InP versus minimum required data rate.}
	\label{Table: Cost of the InP}
	\scalebox{0.99}{
	\begin{tabular}{|cc|c|c|c|c|c|c|c|c|c|}
		\rowcolor[HTML]{38FFF8}
		\hline
		\multicolumn{2}{|c|}{Minimum required data rate (bps/Hz)}                 & 1   & 1.2 & 1.4 & 1.6 & 1.8 & 2   & 3   & 4   & 5   \\ \hline
		\multicolumn{1}{|c|}{\multirow{5}{*}{Cost of the InP (\$)}} & RDPG        & 800 & 770 & 740 & 710 & 690 & 660 & 650 & 650 & 650 \\ \cline{2-11} 
		\multicolumn{1}{|c|}{}                                      & SAC         & 290 & 280 & 270 & 260 & 250 & 240 & 230 & 220 & 220 \\ \cline{2-11} 
		\multicolumn{1}{|c|}{}                                      & DDPG        & 380 & 370 & 360 & 350 & 340 & 330 & 320 & 310 & 310 \\ \cline{2-11} 
		\multicolumn{1}{|c|}{}                                      & Distributed & 310 & 300 & 290 & 280 & 270 & 260 & 250 & 250 & 250 \\ \cline{2-11} 
		\multicolumn{1}{|c|}{}                                      & Greedy      & 180 & 170 & 160 & 150 & 140 & 130 & 120 & 120 & 120 \\ \hline
	\end{tabular}
}
\end{table*}
\subsection{Comparison of Signaling Overhead in the Main Approach and the Baselines}
In all four employed DRL methods to perform actions, the agents require information like the status and received rewards. Therefore, this information must be intercommunicated between the E2E orchestrator, RAN, and core domains. In the centralized approaches we use (i.e., RDPG, SAC, and DDPG), all the information is concentrated in one place, which causes the signaling overhead to increase. But in the distributed method, part of the information is dissolved in the RAN domain, and another part of the information is dissolved in the core domain, so the signaling overhead of the distributed way is less than the centralized way. However, the distributed approach is less performance. To calculate and analyze this information, we consider that each element of the matrices of the channel gain, node, and link can be decoded as a fixed-length 16-bit binary string. Accordingly, we use a type \textit{'float16'} in \textit{Python's Numpy}\footnote{\textit{NumPy} is a library for the \textit{Python} programming language used to perform an extensive range of mathematical operations on arrays.} library. In the RAN domain in our proposed system model with the $I$ BSs, $C$ users, and $K$ subchannels, the total signaling overhead in each episode is equal to:
\begin{align}\label{signal_RAN}
	16 \times I \times C \times K\ \text{bits}
\end{align}
Moreover, in the core domain with the $N$ nodes, $V_{\text{Total}}$ VMs, and $L$ links, the total signaling overhead in each episode is equal to:
\begin{align}\label{signal_core}
	16 \times \left(\left(N\times V_{\text{Total}}\right)+L\right)\ \text{bits}
\end{align}
Therefore, in the distributed method, the amount of signaling overhead in the radio and core parts is equal to formulas \eqref{signal_RAN} and \eqref{signal_core}, respectively. But in the centralized approaches, signaling overhead is equal to the summation of formulas \eqref{signal_RAN} and \eqref{signal_core}. In the end, to better understand the subject, we compare the total signaling overhead in each episode for the RDPG approach and the baselines in Table.\ref{Signaling Overhead}.
\begin{table}[h!]
	\centering
	\caption{Total signal overhead of the algorithms}
	\label{Signaling Overhead}
	\scalebox{0.99}{	
		\begin{tabular}{|c|cc|}
			\hline
					\rowcolor[HTML]{38FFF8} 
			\textbf{Algorithm}           & \multicolumn{2}{c|}{\textbf{Total signaling overhead}}                                                                   \\ \hline
			RDPG                         & \multicolumn{2}{c|}{$16 \times I \times C \times K+16 \times \left(\left(N\times V_{\text{Total}}\right)+L\right)\ \text{bits}$}     \\ \hline
			SAC                          & \multicolumn{2}{c|}{$16 \times I \times C \times K+16 \times \left(\left(N\times V_{\text{Total}}\right)+L\right)\ \text{bits}$}     \\ \hline
			DDPG                         & \multicolumn{2}{c|}{$16 \times I \times C \times K+16 \times \left(\left(N\times V_{\text{Total}}\right)+L\right)\ \text{bits}$}     \\ \hline
			\multirow{2}{*}{Distributed} & \multicolumn{1}{c|}{RAN domain}                      & Core domain                                                       \\ \cline{2-3} 
			& \multicolumn{1}{c|}{$16 \times I \times C \times K\ \text{bits}$} & $16 \times \left(\left(N\times V_{\text{Total}}\right)+L\right)\ \text{bits}$ \\ \hline
		\end{tabular}
	}
\end{table}
\section{Conclusion}\label{VIII}
We studied a resource allocation problem in E2E NwS based on the SDN and NFV concepts by considering uncertainties in the number of slice requests from users, data rate requests in each slice, and CSI. We formulated the utility function of the InP as the difference between revenue and cost in the network architecture. The proposed problem was formulated as non-convex mixed-integer non-linear programming. Due to the complexity of the problem and many actions and states space, we employed several DRL algorithms. Because of the uncertainties in the problem, we considered the RDPG method as the main solution and compared it with other methods under various aspects. According to the simulation results, the SAC method is better than the DDPG, distributed, and greedy approaches, respectively. In addition, the RDPG strategy outperforms the SAC approach by, on average, 70$\%$. Therefore, the RDPG method is robust for our proposed problem and is considered as the main method.
\bibliographystyle{ieeetr}
\bibliography{Cite}
\end{document}